\documentclass[usenatbib,useAMS]{mn2e}
\usepackage{epsfig}
\usepackage{times}
\usepackage{amsmath}

\usepackage{verbatim}

\def\d{\mbox{d}}
\newcommand{\simgt}%
        {\,\hbox{\lower0.6ex\hbox{$\sim$}\llap{\raise0.6ex\hbox{$>$}}}\,}
\newcommand{\simlt}%
        {\,\hbox{\lower0.6ex\hbox{$\sim$}\llap{\raise0.6ex\hbox{$<$}}}\,}

\title[]{Multi-dimensional modelling of X-ray spectra for 
AGN accretion-disk outflows II}
\author[Sim et al.]{S. A. Sim$^1$, L. Miller$^2$, K. S. Long$^3$,
  T. J. Turner$^{4,5}$, J. N. Reeves$^6$\\
$^{1}$Max-Planck-Institut f\"{u}r Astrophysik,
Karl-Schwarzschildstr. 1, 85748 Garching, Germany\\
$^{2}$Dept. of Physics, University of Oxford, Denys Wilkinson
Buiding, Keble Road, Oxford OX1 3RH, U.K.\\
$^{3}$Space Telescope Science Institute, 3700 San Martin Drive,
Baltimore, MD 21218, U.S.A\\
$^{4}$Dept. of Physics, University of Maryland Baltimore County, 1000 Hilltop Circle, Baltimore, MD 21250, U.S.A\\
$^{5}$Astrophysics Science Division,
NASA/GSFC, Greenbelt, MD 20771, U.S.A\\
$^{6}$Astrophysics Group, School of Physical and Geographical
Sciences, Keele University, Keele, Staffordshire ST5 8EH, U.K.\\
}

\date{\today}

\pagerange{\pageref{firstpage}--\pageref{lastpage}}
\pubyear{}
\begin{document}
\maketitle
\label{firstpage}

\begin{abstract}
Highly-ionized fast accretion-disk winds have been suggested as an
explanation for a variety of observed absorption and emission features
in the X-ray spectra of Active Galactic Nuclei. Simple estimates have
suggested that these flows may be massive enough to carry away a significant
fraction of the accretion energy and 
could be involved in creating the link between supermassive
black holes and their host galaxies. However, testing these hypotheses,
and quantifying the outflow signatures, requires high-quality
theoretical spectra for comparison with observations. Here we describe
extensions of our
Monte Carlo radiative transfer code that allow us to generate
realistic theoretical spectra for a much wider variety of disk wind models 
than possible in our previous work. In particular, we
have expanded the range of atomic physics simulated by the code 
so that L- and M-shell ions can now be included.
We have also substantially improved our treatment of both ionization and
radiative
heating 
such that
we are now able to compute spectra for outflows containing far more
diverse plasma conditions. 
We present example calculations
that illustrate the variety of spectral features predicted by
parameterized 
outflow models and demonstrate their applicability to the
interpretation of data by comparison with observations of the bright
quasar PG1211+143. We find that the major features in the observed
2 -- 10~keV spectrum of this object can be well-reproduced by our
spectra, confirming that it likely hosts a massive outflow. 
\end{abstract}


\section{Introduction}
\label{sect_intro}

The vast
energy released in accretion by a supermassive black hole means that
active galactic nuclei (AGN) could have a fundamental role in
the formation and evolution of galaxies (see \citealt{cattaneo09}). This
depends however on the nature of the physical mechanisms that
couple the AGN to its host galaxy. One such possible AGN feedback
mechanism is that of a massive outflow, launched somewhere close to
the accreting black hole. It has already been shown that such flows
might be able to account for observed correlations between properties
of nuclear black holes and their host galaxies (see e.g. \citealt{king03b,king05}).

Direct evidence for AGN outflows and quantification of their
properties, however, requires interpretation of observational
data. Perhaps the best evidence for energetically important outflows
around AGN comes from the detection of X-ray absorption lines identified
with significantly blueshifted line transitions of highly ionized
material (see \citealt{turner09} for a recent review of
X-ray observations of AGN). The most promising origin for mass-loss in
the immediate vicinity of an accreting supermassive black hole is an
accretion disk wind. Hydrodynamical simulations have demonstrated
that such flows are plausible (e.g. \citealt{proga04}). However, these 
simulations also suggest that disk winds are likely to be
very complex and thus to interpret their spectroscopic signatures
quantitatively requires detailed synthetic spectra for comparison.

In previous work (\citealt{sim05b}, \citealt{sim08}, hereafter
Paper~I), we developed
a numerical code for computing synthetic X-ray spectra for outflow models.
We showed that simply-parameterized disk wind models could readily
account for strong
blueshifted absorption features associated with Fe~{\sc xxv} and
Fe~{\sc xxvi} and we explored some of the effects of the outflow
density, geometry and ionization state on these absorption line properties.
Our models also confirmed that a very highly ionized outflow could 
affect the X-ray spectrum in more subtle ways. In particular, the blueshifted
Fe absorption lines have associated emission features that are broad
and can develop extended red-skewed wings owing to the effects of
electron scattering in the flow (see also
\citealt{laming04,laurent07}). We concluded that, although complex,
outflow models have the promise to simultaneously 
account for several of the observed X-ray spectroscopic features of
AGN. We illustrated this via a direct comparison to observations
of the well-known narrow line Seyfert 1 galaxy Mrk~766.
However, that work was limited to the treatment of only the atomic
physics necessary for the most highly ionized atomic species
(specifically, He- and H-like ions). Although AGN provide copious
ionizing radiation, 
the region responsible for the primary
X-ray emission is assumed to be small and centrally concentrated. Therefore
more distant portions of an outflow -- or those shielded from the
X-ray source by other material -- need not be so significantly
ionized. Thus a more realistic study requires that a wider range of
ionization conditions can be considered. In particular, photoelectric
absorption and/or fluorescent
emission by lower ionization states of Fe may affect the X-ray
spectrum, both in the Fe~K region and at lower X-ray energies. 
For lower typical ionization conditions, the relative importance of
the lighter elements also grows significantly and they can lead to strong
photoelectric absorption for energies $\sim 1$~keV and associated
discreet line features in the soft X-ray spectrum as have been
reported in observations
(e.g. \citealt{pounds03,pounds05,pounds07,pounds09}).

Here, we extend our Monte Carlo radiative transfer code
described in Paper~I to treat the physics of L- and M-shell ions of 
astrophysically abundant elements. We also substantially improve the
treatment of the ionization state and introduce a simple means of
estimating the kinetic temperature in the outflow, thereby eliminating
this as a free parameter of the models. In addition, we modify the
means by which the orientation-dependent 
spectra are extracted from our simulations in order to
suppress the level of Monte Carlo noise.
These improvements to the code
and atomic data used are described in Section~\ref{sect:method}. In
Section~\ref{sect:iontest} we demonstrate that our improved treatment
of ionization leads to quantitatively good agreement with well-known 1D
codes and in Section~\ref{sect_example} we discuss results from a
calculation made with the improved code. Finally, in
Section~\ref{sect_pg12}, we discuss the comparison of spectra computed
with our models to observations of the quasar PG1211+143 before
drawing conclusions in Section~\ref{sect_conc}.

\section{Method}
\label{sect:method}

\subsection{Code overview}

The code operates by performing a sequence of Monte Carlo radiative
transfer simulations employing an indivisible packet scheme (see
e.g. \citealt{lucy02,lucy03}). In each simulation, we follow the
propagation of Monte Carlo
quanta (``$r$-packets'') representing bundles of X-ray photons through
simply-parameterized models for accretion disk winds. During the
$r$-packet propagation,
the effects of Compton scattering, photoelectric absorption,
bound-bound line interactions (for which the Sobolev approximation is adopted) 
and free-free absorption are simulated in detail. Interactions
with matter (specifically bound-bound and bound-free transitions) are
treated using the Macro Atom
formalism introduced by \citet{lucy02}. This 
approach allows for a full treatment of line scattering, recombination
and fluorescence as required to produce realistic spectra for
comparison with observations. The code has also been extended to use
the 
method of \citet{lucy02,lucy03} for treating both the
radiative heating and cooling of matter.
In this approach, when the MC quanta
undergo physical events in which the energy they represent is used to heat the
plasma (e.g. by free-free absorption of photons), they are instantaneously
converted to packets of thermal kinetic energy (so-called $k$-packets).
Assuming thermal
equilibrium, these $k$-packets are then eliminated by randomly sampling the
available cooling processes (see Section~\ref{section:tb}). Depending on the cooling process
selected, the $k$-packet may be converted to an $r$-packet which is
then free to propagate. The $r$-packets created in this way 
simulate the cooling radiation emitted
by the wind.

During each Monte Carlo simulation, the physical
wind properties (e.g. temperature, ionization conditions etc.) are
held fixed. At the end of each such simulation, the histories of the Monte
Carlo quanta are used to make improved estimates of the wind
properties assuming radiative, thermal and ionization equilibrium and
then the Monte Carlo experiment is repeated. In this way,
the wind properties are iterated to consistency with the radiation
field. After the iteration cycle, a final Monte Carlo simulation is
performed from which viewing-angle dependent spectra are extracted
(see Section~\ref{section:extract}).

\subsection{Ionization}

To provide a more complete description of the ionization state of the
outflow, the modified nebular approximation adopted in Paper~I has
been replaced with a detailed computation of ionization and
recombination rates that are then used to solve for the steady-state
ionization fractions. 
The following sections describe the processes
that are included.

\subsubsection{Photoionization}
\label{sect:Gamma}

Bound-free absorption involving the ejection of an outer shell
electron is included for ground states and low-energy metastable
levels of all ions. 
During the Monte Carlo simulation, the packet
trajectories are used to record estimators for the photoionization
rate coefficient ($\gamma$) for each bound-free process in each of the
computational grid cells of wind properties. The estimator for the
rate coefficient of bound-free absorption from level $j$ in a cell $p$ is
given by

\begin{equation}
\gamma_{j,k} = \frac{1}{V_{p} \; \Delta t}\sum_{\nu > \nu_0} \frac{a_{j}(\nu)}{h
  \nu} \epsilon \, \d s
\end{equation}
where $V_{p}$ is the wind cell volume, $a_{j}$ is the photoabsorption
cross-section for level $j$ (see Section~\ref{sect_atomicdata}) and the
summation runs over all Monte Carlo quanta trajectories inside cell
$p$ when the quanta have co-moving frame frequency $\nu$ that is greater than the
threshold $\nu_0$. $\epsilon$ is the co-moving frame energy of the
quantum and $\d s$ is the trajectory length. 
$\Delta t$ is the time interval represented
by the Monte Carlo experiment. The Doppler factor correction term for
transforming the photon path length from the observer frame to the
fluid frame (see e.g. \citealt{lucy05}) is 
neglected for computational expediency.

At the end of a Monte Carlo simulation, the $\gamma$
estimators for levels of each ion are combined to make a
photoionization rate estimator for each ion ($I$) in each cell ($p$) defined by

\begin{equation}
\Gamma_{I,k} = \sum_{j \in I} n_{j,k} \gamma_{j,k} / \sum_{j \in I} n_{j,k}
\end{equation}
where the summations run over all level $j$ of the ion
for which bound-free absorption has been included in the Monte Carlo
simulation. 
The level populations, $n_{j,k}$ are discussed in Section~\ref{sect:excitation}.

For L- and M-shell ions, photoabsorption involving inner 
shell electrons is also included. In the treatment of these
processes, details of the outer electron
configuration are neglected and all levels of the ion are assigned the
same cross section. 
Following inner shell photoabsorption, the newly made
ion is generally in a highly-excited state that may either decay
radiatively or via ejection of one or more Auger electrons. 

In the case of K-shell photoionization of L-shell ions, at most one Auger electron is expected to
be produced -- thus the ionization state may ultimately increase by either one or two.
To incorporate this in the ionization
balance, Monte Carlo estimators are used to obtain a rate
coefficient for K-shell photoabsorption for each ion

\begin{equation}
\Gamma^K_{I,k} = \frac{1}{V_{p} \; \Delta t}\sum_{\nu > \nu_K} \frac{a^{K}_{I}(\nu)}{h
  \nu} \epsilon \, \d s
\end{equation}
where the summation now runs over all packets having frequency about
the K-shell edge threshold, $\nu_K$ and $a^{K}$ is the K-shell
photoabsorption cross-section. To obtain separate rates for single and
double ionization, the mean probability ($p_{I}[1]$) of K-shell absorption being
followed by ejection of a single Auger electron is required. This is computed using

\begin{equation}
p_{I}[1] = < \frac{\sum_i A^{a}_{ji}}{\sum_i A_{ji}\beta_{ji} + \sum_i
  A^{a}_{ji}} >_{j}
\end{equation}
where $< ... >_{j}$ is a weighted mean over all
states $j$ of the ionization state $I+1$ that are accessible by
K-shell photoabsorption of ion $I$, $\sum_i A^a_{ji}$ is the total autoionization
rate out of level $j$ and $\sum_i A_{ji} \beta_{ji}$ is the sum of
radiative decay rates incorporating the Sobolev escape probabilities
$\beta_{ji}$ over all downward transitions from level $j$. In
practice, only states reached by K-shell photoabsorption from the
ground configuration of ion $I$ are included in the average and these
are weighted with their statistical weights. 
The estimated rate at
which K-shell photoabsorption in an L-shell ion increases the ionization state by one is
then given by

\begin{equation}
\Gamma^K_{I,k} (1 - p_{I}[1])
\end{equation}
and, for double ionization,

\begin{equation}
\Gamma^K_{I,k} p_{I}[1] \; \; \; .
\end{equation}
Data sources for $A_{ji}$ and $A^{a}_{ji}$ are discussed in Section~\ref{sect_atomicdata}.

K- and L-shell photoionization of M-shell ions, are treated in a
similar manner, recording a $\Gamma^{K/L}$ estimator for each
such process in each grid cell. Owing to the complexity of the
subsequent decays of the vacancy state (which lead to the ejection of
up to three Auger electrons for K-shell ionization of the M-shell ions
we consider),
the Auger ejection probabilities $p_{I}[1]$, $p_{I}[2]$ and
$p_{I}[3]$, are not computed from the atomic data set used by the code but taken from
the yields tabulated by \cite{kaastra93}. This approach obviates 
the need to follow the decay chains of the vacancy states in
detail but comes at the expense of assuming that all the relevant
radiative transitions are optically thin such that the M-shell ion populations
are unaffected by photon trapping.

\subsubsection{Collisional ionization and recombination processes}

In addition to the photoionization and Auger processes described
above, the ionization balance includes collisional ionization, both
direct ($C_{DI}$) and due to collisional excitation followed by
autoionization ($C_{EA}$), and both radiative ($\alpha^{r}$) and di-electronic
($\alpha^{d}$) recombination (see Section~\ref{sect_atomicdata} for
data sources).

\subsubsection{Ionization equilibrium}

At the end of each Monte Carlo simulation, the computed
photoionization rate coefficients ($\Gamma$; see
Section~\ref{sect:Gamma}) are used to solve for an improved set of ion populations
for each element, assuming ionization equilibrium and including all the
processes described above. Since the recombination and collisional
ionization terms depend on the adopted kinetic temperature ($T_{e}$), the
ionization balance is iterated to consistency with the calculation of
$T_{e}$ (see below) for fixed
Monte Carlo estimators.

\subsection{Thermal balance}
\label{section:tb}

In Paper~I, $T_{e}$ was assumed to be uniform and
treated as a model parameter. In order to address more general and
realistic wind conditions, here we relax this assumption and obtain
an estimate of $T_{e}$ as a function of position via a simplified
treatment of the heating and cooling rates in the wind and the
assumption of thermal equilibrium. 

\subsubsection{Heating rates}

The treatment includes heating by
Compton scattering, free-free absorption and bound-free absorption of
the X-ray radiation field described by the Monte Carlo
simulations. Estimators for each of these heating processes in each
grid cell can be readily constructed following \citet{lucy03}. For
Compton scattering, the heating rate in a grid cell ($p$) is given by

\begin{equation}
H^C_p = \frac{n_{e}}{V_{p} \; \Delta t} \sum \bar{f}(\nu) \, \sigma(\nu) \,
  \epsilon \, \d s
\end{equation}
where the summation runs over all packet trajectories within the grid
cell $p$, $\sigma(\nu)$
is the Compton cross-section for the frequency ($\nu$) of the packet
and $\bar{f}(\nu)$ is the mean energy lost per Compton scattering
event. The free-free heating rate is

\begin{equation}
H^{ff}_p = \frac{1}{V_{p} \; \Delta t} \sum \kappa_{ff}(\nu) \,
  \epsilon \, \d s \; \; .
\end{equation}
Again, the summation runs over all packet trajectories in the cell.
The
free-free absorption coefficient is given by

\begin{equation}
\kappa_{ff}(\nu) = 3.69 \times 10^8 \nu^{-3} T_{e}^{-1/2} n_{e} (1 -
e^{-h\nu/kT_e}) \sum_{I} Z_{I}^2 N_{I} \; \; \mbox{cm$^{-1}$}
\label{eqn:ffop}
\end{equation}
where the summation runs over all ions ($I$) and $Z_{I}$ is the ion
charge.

Heating rates for photoabsorption are obtained in a similar fashion to
the ionization rate estimators in \ref{sect:Gamma}. For bound-free
absorption involving outer shells, level-by-level estimators are
recorded and used to obtain a total heating rate for each ion $I$, 

\begin{equation}
H^{bf}_{I,k} = \sum_{j \in I} n_{j,k} h^{bf}_{j,k}
\end{equation}
where

\begin{equation}
h^{bf}_{j,k} = \frac{1}{V_{p} \; \Delta t}\sum_{\nu > \nu_0}
{a_{j}(\nu)} \left({1 - \frac{\nu_{0}}{\nu}}\right) \epsilon \, \d s \; \;.
\end{equation}
Similarly, the heating rate due to K-shell photoabsorption by ion $I$ is given by

\begin{equation}
H^{K}_{I,k} =  \frac{N_{I}}{V_{p} \; \Delta t}\sum_{\nu > \nu_K}
{a^{K}_{I}(\nu)} \left({1 - \frac{\nu_{K}}{\nu}}\right) \epsilon \, \d s \; \;.
\end{equation}

All other possible heating sources are neglected, including any contributions
from outside the spectral region of the simulated radiation field or
by any non-radiative processes.

\subsubsection{Cooling rates}
\label{sect:cooling}

Cooling rates are required for both the treatment of $k$-packets (see
Section~\ref{sect:kpkt}) and to obtain an estimate of the local kinetic temperature.
Cooling due to radiative recombination, electron
collisional excitation, free-free emission, Compton cooling by
low-energy photons and the adiabatic expansion of the outflow are
included in this calculation.

 The cooling rates due to spontaneous radiative recombination
$C^{bf}_{I,k}$ and electron collisional excitation $C^{cl}_{I,k}$ are obtained
following exactly Lucy (2003; his equations 31 and 33).
The free-free cooling rate is 

\begin{equation}
C^{ff}_{p} = 1.426 \times 10^{-27} {T_{e}^{1/2}} n_{e} \sum_{I} Z_{I}^2
N_{I} \; \mbox{ergs~cm$^{-3}$~s$^{-1}$}
\end{equation}
where we follow \citet{lucy03} in setting the mean Gaunt factor to
one. The adiabatic
rate cooling rate is approximated by

\begin{equation}
C^{a}_{p} = 1.5 n_{g} k T_{e} {\mbox{\boldmath$\nabla\cdot v$}}
\end{equation}
where $n_{g}$ is the total particle density and {\boldmath$\nabla\cdot
  v$} is the divergence of the velocity, evaluated at the midpoint of the cell.

The cooling rate for a population of non-relativistic electrons due to Compton up-scattering of low-energy photons ($h
\nu << k T_{e}$) can be estimated from the local energy density of such
photons $U_{\gamma}$ via

\begin{equation}
C^C_{p} = 4 \sigma_{T} U_{\gamma} \frac{k T_{e}}{m_{e} c}
\end{equation}
where $\sigma_T$ is the Thomson cross-section (see \cite{frankbook}, p. 180).
The cooling rate due to Compton up-scattering of photons is expected to
be dominated by interactions with the low energy photons 
originating from the accretion disk since they should significantly
outnumber the X-ray photons for which our detailed radiative transfer
calculations are performed. An accurate treatment of Compton cooling
would therefore require that the entire bolometric light output of
the system be considered. However, since our primary concern is only
to obtain a reasonable estimate of the kinetic temperature, we
avoid this complication and derive an estimate for the local value of 
$U_{\gamma}$ from the total bolometric luminosity of the source
($L_{\mbox{\scriptsize bol}}$, which is treated as an input parameter
for the model) and an assumption of spherical
geometric dilution, i.e.

\begin{equation}
U_{\gamma} = \frac{L_{\mbox{\scriptsize bol}}}{4 \pi r^2 c}
\label{eqn:compcool}
\end{equation}
where $r$ is the distance from the centre of the grid cell to the
coordinate origin. This expression is only a crude estimate for $U_{\gamma}$
since it neglects both the geometry of the emission
regions for low-energy photons (i.e. it assumes a centrally
concentrated point source) and the opacity of the
outflow to low-energy photons. Nevertheless, it provides a convenient
estimate of the Compton cooling rate 
that introduces no additional computational demands 
and should be reasonably accurate, at least for the
inner (hottest) regions of the wind.

\subsubsection{The kinetic temperature determination}
\label{sect:get_tb}

To obtain the kinetic temperature ($T_{e}$) for each grid cell, the
radiative heating rates computed from the Monte Carlo simulation are
compared with cooling rates to estimate the temperature at which
thermal equilibrium,

\begin{equation}
H^C_p + H^{ff}_p + \sum_{I} (H^{bf}_{I,k} + H^{K}_{I,k})
= C^{ff}_{p} + C^{a}_{p} + C^{C}_p + \sum_{I} (C^{bf}_{I,k} + C^{cl}_{I,k})
\label{eqn:tb}
\end{equation}
 is established.
In practice, the thermal balance equation (equation~\ref{eqn:tb}) is solved
iteratively with the ionization balance equations to obtain
self-consistent estimates of both the
ionization fractions and $T_{e}$ for each grid cell.

\subsubsection{Excitation}
\label{sect:excitation}

In principle, the excitation state should be obtained from the 
statistical equilibrium equations. However, to do so would require 
computation and storage of
estimators for all bound-bound transitions in every cell which is
prohibitive for large computational grids. Therefore, we adopt a very
simplified treatment of excitation, namely we use the Boltzmann
distribution at the local kinetic temperature

\begin{equation}
\frac{n_{i}}{n_{\mbox{\scriptsize g.s.}}} = \frac{g_{i}}{g_{\mbox{\scriptsize g.s.}}} \exp(-\epsilon_{i}/kT_{e})
\end{equation}
where $n_{i}$ is the population of a state with statistical weight
$g_{i}$ and energy $\epsilon_{i}$, relative to the grounds state (g.s.).
For the wind models we will describe in Sections~\ref{sect_example} and \ref{sect_pg12},
our simplified treatment of excitation is not expected to
significantly affect the Fe K$\alpha$ region of the spectrum since the line
features which form there are mostly associated with low excitation
states of high-ionization state material. The soft regions ($\simlt
1$~keV) are more likely to be affected owing to the large numbers of
transitions between excited states of L-shell Fe and Ni ions which
contribute to the opacity at these photon energies.

\subsection{Monte Carlo simulations}

The radiation transport simulations are performed using the
scheme described in Sections 3.1 to 3.3 of Paper~I, modified as
described below.

\subsubsection{Initialization of packets}

As in Paper~I, an initial power-law spectrum of packet energies is adopted but
the input spectrum now extends from 0.1~keV up to 511~keV to allow
comparison with
observational constraints from instruments with significant effective
area at relatively high photon energies.

To account for the effects of a small but non-zero angular size of the
X-ray emission region, the packets are no longer initialized exactly
at the coordinate origin but in a spherical region around the
origin. The radial extent of the emission region, $r_{er}$, is a
parameter in the model and is generally chosen to be several
gravitational radii, as appropriate for an X-ray
emission region of size comparable to the innermost
radii of an accretion disk.

\subsubsection{Propagation of packets}

Compton scattering, bound-free absorption and line absorption are
treated exactly as described in Paper~I, except that bound-free
absorption by a few low-lying metastable levels is included for most
ions. 

Free-free opacity from all ions is included using
equation~\ref{eqn:ffop}. Following free-free absorption, packets are
always converted to thermal energy ($k$-packets; see \citealt{lucy03}).

K-shell photoionization is included as an opacity source for
all L-shell and M-shell ions. Details of the outer electron
configuration are neglected such that all states of an ion contribute
to the same opacity term. When K-shell photoabsorption occurs in the
simulations, the packet either activates a macro atom or is converted
to a $k$-packet -- this is exactly analogous to the
 treatment of outer-shell bound-free processes which are discussed in
 detail by \cite{lucy03}. Here, the probability of $k$-packet
 conversion is simply given by

\begin{equation}
p_{k} = 1 - \nu_K/\nu
\end{equation}
where $\nu$ is the absorption frequency and $\nu_K$ is the
relevant K-shell absorption edge frequency.
If the outcome of K-shell photoionization is activation of a macro
atom, the state activated must have a K-shell vacancy in its
configuration and will
generally be highly excited \footnote{Conversely,
following activation by outer-shell bound-free absorption, macro atoms
are always assumed to be in the ground
state of the newly formed ion.}. 
Which of the K-shell vacancy
states in the newly formed ion is activated is chosen randomly by
sampling the statistical weights of the viable states in the
atomic data set.
 
\subsubsection{Macro Atom processes}

The macro atom treatment used in Paper~I has been extended to
incorporate autoionization, bound-free transitions from excited states and K-shell
photoionization.

Autoionization allows activated macro atom states to de-excite by
conversion to a $k$-packet and to make internal transition to higher
ionization states. Following the argumentation of \cite{lucy02}, the
deactivation probability (macro atom $\rightarrow$ $k$-packet)

\begin{equation}
p^D = N_{i} A_{a} (\epsilon_i - \epsilon_f)
\end{equation}
where $A_{a}$ is the rate coefficient ([s$^{-1}$]) for
autoionization from initial state $i$ to final state $f$,
$\epsilon_i$ and $\epsilon_f$ are the energies (excitation
plus ionization) of the states, and $N_{i}$ is the
normalization factor. Similarly, the
probability of making an internal macro atom jump $i \rightarrow f$ is

\begin{equation}
p^J_{i \rightarrow f} = N_{i} A_a \epsilon_f \; \; .
\end{equation}
This formulation requires that the autoionization process
connects two well-defined states in the atomic data set. In our
calculations, this is done for L-shell ions using level-to-level
autoionization rates (see Section~\ref{sect_atomicdata} for data
sources). 
However, for the M-shell ions
we do not consider the target level in detail but assume
that the energy flow in the autoionization process is dominated by the
energy carried by the Auger
electron -- i.e. we assume that $\epsilon_f << \epsilon_i$ such that
$p^J/p^D \approx 0$. This has the advantage that we do not require
atomic level-to-level autoionization rates but only total Auger-widths
for vacancy states, as are available from the literature (see Section~{\ref{sect_atomicdata}).

For L-shell ions di-electronic recombination, is also
included in the macro atom scheme. 
Di-electronic rate coefficients, $\alpha_d$, are obtained
from $A_a$ in the usual manner and used to formulate an
internal transition probability between the states

\begin{equation}
p^{J}_{i \rightarrow f} = N_{f} \alpha_d \epsilon_f n_{e}
\end{equation}
where $n_{e}$ is the local electron number density. Note that, as for
e.g. bound-free absorption, there is no macro-atom deactivation
probability associated with $\alpha_d$.

The $\gamma$ estimators described in Section~\ref{sect:Gamma} are used
to compute bound-free macro-atom internal transition probabilities
following \cite{lucy03}. In all such cases, it is assumed that the
relevant upper state is the ground state of the target ion. Similarly,
the $\gamma^{K}$ estimators are used to include internal transitions
associated with inner shell ionization. These are applied to all
energy levels of the absorbing ion lying below the first ionization
potential and connect to K-vacancy states of the next higher
ionization state. The choice of which K-vacancy state to activate
following an internal transition is made
by randomly sampling their statistical weights.
L-shell ionization of M-shell ions is not included in the macro atom treatment.

\subsection{Accretion disk}
\label{sect:disk}

We assume that the black hole is surrounded by an optically thick
accretion disk that lies in the $xy$-plane and extends from an inner
radius $r_{d}$ outwards throughout the simulation domain. $r_{d}$ is
treated as a model parameter but set equal to 6~$r_{g}$ (where $r_{g} = G M_{bh} / c^2$) for all the
simulations performed here.
During the Monte Carlo simulations, we also assume that all photon
packets that strike the disk are lost to the X-ray regime and they
are removed from the simulation. 

For our particular choice of source
geometry, the disk subtends a modest solid angle as seen
by the X-ray source ($\sim 2.8$~sr), 
meaning that a fraction of the primary X-ray photons strike the inner
regions of the disk
directly, without interaction in the wind.\footnote{This fraction, however, is
not physically meaningful in our model since it is merely
determined by our simple choice for the geometry of the X-ray
emitting region.}
The total flux striking the disk, however, is greater than this since
the wind effectively causes irradiation  
of the outer parts of the disk
via scattering and emission in the wind.
Physically, these photons will interact
with the disk and may lead to an additional component of disk
reflection in the X-ray spectrum. We neglect any such component here
but it will be investigated in a future study.

\subsection{Thermal emission and $k$-packets}
\label{sect:kpkt} 

In paper~I, it was assumed that energy packets converted to
$k$-packets would be lost to the hard X-ray band. However, since
cooling rates are now computed in order to impose thermal equilibrium (Section~\ref{sect:cooling}),
these are now used to re-emit the packet, thus simulating the thermal emission of the outflow.

When conversion to a $k$-packet occurs, 
the cooling terms described in Section~\ref{sect:cooling} are first
randomly sampled to choose a cooling process. If $C^{a}$ is chosen,
the energy is assumed to be lost to the radiation field. 
The collisional cooling processes ($C^{cl}$) lead to activation of a
macro atom state, the particular state being chosen by sampling the
terms contributing to $C^{cl}$ (see \citealt{lucy03}). The full macro
atom machinery then determines the outcome of this event.

Choosing $C^{fb}$ or
$C^{ff}$ leads to conversion to photon packets. Both
processes are assumed to emit isotropically in the co-moving frame and
the emission frequency is selected using equations 26 and 41 of
\cite{lucy03}. Note that this emission rule for bound-free processes
is not exact here since it assumes a simplified cross-section shape
but it is significantly less computationally demanding that sampling
the real emissivity and is adequate for current purposes.

\subsection{Extraction of spectra}
\label{section:extract}

Although spectra may be obtained by directly binning the emergent
packets by frequency and direction of travel, as done in Paper~I, this
has the disadvantage that very large numbers of quanta must be
simulated to suppress Monte Carlo noise, particularly when the spectrum
is expected to depend strongly on the observer line-of-sight. As
discussed by e.g. \cite{lucy99}, that approach is far
from optimal and does not make full use of the packet trajectories
during the Monte Carlo simulation. 
Ideally, volume-based Monte Carlo estimators should be used to obtain
emissivities throughout the volume of the simulation 
and used to obtain spectra via a 
formal solution of the radiative transfer equation
(see discussion e.g. \citealt{lucy99}). However, this approach is very
demanding of memory resources: for our atomic data set (with $\> 10^5$ atomic processes) and 
computational
grids (100$\times$100 wind cells), it ideally requires the storage of $\sim
10^9$ floating-point estimators which is prohibitive. Therefore we
extract spectra using a approach similar to that implemented in the
disk-wind simulations of \cite{long02} which, although less efficient,
does not make significant demands on memory consumption.

Viewing-angle dependent spectra are only extracted during the final
Monte Carlo simulation, once the ionization and thermal properties of
the outflow have been iterated to consistency with the radiation
field. The set of lines-of-sight for which spectra are required is
pre-specified and then every {\it physical} event (meaning emission or
scattering of a photon-packet) that occurs in the
Monte Carlo simulations is used to compute a contribution to each of
the spectra in the following manner.

When a physical event takes place in the Monte Carlo simulation, the
propagation of the packet is temporarily suspended and the probability
per unit solid angle
of the the associated photon-packet having been scattered into each of
the requested lines-of-sight is computed, $\d P(${\boldmath
  $n$}$)/\d \Omega$ where the vector {\boldmath $n$} identifies the
line-of-sight. Depending on the physical process responsible for the
packet emission/scattering, $\d P(${\boldmath
  $n$}$)/\d \Omega$ may depend on previous properties of the packet
(e.g. its incoming direction in the case of Compton scattering) or on
the local outflow properties (e.g. the Sobolev escape probability in
the case of line emission). We then integrate the total optical depth,
$\tau(${\boldmath $n$}$)$,
from the position of the physical event to the edge of the wind in the
direction {\boldmath $n$}. $\tau(${\boldmath $n$}$)$ is computed
exactly as in the Monte Carlo simulation,
incorporating contributions from all physical processes included in
the code and allowing for Doppler shifts to bring the photons into
resonance with spectral lines. With $\d P(${\boldmath
  $n$}$)/\d \Omega$ and $\tau(${\boldmath $n$}$)$ computed, we then add
an energy contribution from the event to the spectrum associated with
direction {\boldmath $n$} of

\begin{equation}
\epsilon_{rf} \frac{\d P(\mbox{\boldmath $n$})}{\d \Omega} e^{-\tau(\mbox{\boldmath $n$})}
\end{equation}
where $\epsilon_{rf}$ is the rest-frame energy of the packet and the
contribution is added to the frequency bin associated with the
rest-frame frequency of the packet. This calculation is repeated for
each of the lines-of-sight for which spectra have been requested and
then the Monte Carlo simulation proceeds until the next physical event
occurs. In this way, we obtain high signal-to-noise spectra without
resorting to coarse angular binning of the emergent spectra or requiring
the storage of prohibitively large numbers of Monte Carlo estimators.

\subsection{Rescaling of model parameters}
\label{sect_rescale}

Although our numerical simulations are performed adopting a
particular value for the mass of the central black hole
($M_{\mbox{\scriptsize bh}}$), the spectra obtained are applicable for
other  $M_{\mbox{\scriptsize bh}}$-values following 
a rescaling of the dimensional wind
parameters. Specifically, the ionization state, optical depths and
velocity-law in the wind are preserved under a global rescaling where
all the model luminosity parameters ($L_{\mbox{\scriptsize bol}}$,
$L_{X}$) scale with the Eddington
luminosity ($L_{\mbox{\scriptsize Edd}}$),
all lengths ($r_{\mbox{\scriptsize min}}$, $r_{\mbox{\scriptsize max}}$,
$d$, $R_{v}$, $r_{d}$ and $r_{\mbox{\scriptsize em}}$) are scaled to
the gravitational radius ($r_{g}$) and the mass-loss rate ($\dot{M}$) is scaled
to the Eddington accretion rate ($\dot{M}_{\mbox{\scriptsize Edd}}$).

\subsection{Atomic data}
\label{sect_atomicdata}

Table~\ref{tab_atoms} lists the elements and ionization stages for
which atomic models are
included in the radiative transfer calculations. Compared to Paper~I,
the L-shell ions of the astrophysically abundant intermediate-mass and
iron group elements have been added (Mg, Si, S, Ar, Ca, Fe and
Ni). The highest M-shell ions (down to the Cl-like ion) have also been added for
Fe and Ni.
The solar element abundances of \citet{asplund05} are adopted.

\begin{table}
\caption{Elements and ions that are included in the radiative transfer
calculations.}
\label{tab_atoms}
\begin{tabular}{ccccc}
Element & Ions &~~~~~~~~ & Element & Ions\\ \hline
C & {\sc iv -- vii} & &S & {\sc viii -- xvii}\\
N & {\sc v -- viii} & &Ar &{\sc x -- xix}\\
O & {\sc vi -- ix} & &Ca &{\sc xii -- xxi}\\
Ne & {\sc viii -- xi} & &Fe & {\sc x -- xxvii}\\
Mg & {\sc iv -- xiii} & &Ni & {\sc xii -- xxix}\\
Si & {\sc vi -- xv} & & \\ \hline
\end{tabular}
\end{table}

In all cases, inner-shell photoionization cross-sections were taken
from the fits by \cite{verner95}. We note that, although convenient,
those data do not account for smoothing of the K edges by broad
resonances below threshold (see
\citealt{palmeri02,kallman04}). Depending on the ionization state, this
effect could be critical for quantitative analysis of the region around the
Fe K edge but it is not important for the He- and H-like ions
which dominant in most of the wind models we describe in
Sections~\ref{sect_example} and \ref{sect_pg12}.
For outer-shell photoionization from
ground configurations (and also low-lying metastable states),
cross-sections were taken from \cite{verner96} and a simple hydrogenic
approximation was used for cross-sections from more highly excited states.

Atomic energy levels, bound-bound $A$-values and electron collision
strengths were extracted from the CHIANTI atomic database, version 5
\citep{dere97, landi06} for all the included L- and M-shell ions of C to
Fe. 
For the M-shell ions of Fe, the data were limited to those of the
configurations included in the study by \cite{mendoza04}. In all
cases, the $T_{e}$ dependence of the electron collision strength was
neglected and the low-temperature limit adopted.

Autoionization rates for K-vacancy states in L-shell ions of Fe
were taken from \cite{palmeri03} and Auger widths for the K-vacancy states of the M-shell Fe
ions from \cite{mendoza04}. 
The total
autoionization rates for the K-vacancy states of Mg, Si, S, Ar and Ca were taken from
\cite{palmeri08} and the branching ratios of the autoionization
process into the specific final states of the target ion were assumed
to be the same for the iso-electronic Fe ion. 

For the included L- and M-shell ions of Ni ({\sc xii -- xix}), energy
levels, Einstein A-values and total Auger widths were taken from
\cite{palmeri08b}. Again, the L-shell autoionization branching ratios were
assumed to be the same as in the iso-electronic Fe ion.

Radiative recombination rates are taken from \cite{gu03}, where
available, or else from \cite{verner96b} (H-, He- and Li-like ions of
C, N, O and Ne and Na-like ions of Fe and Ni),
\cite{arnaudray} (Fe~{\sc xvi}) or \cite{shull82,shull82a}, otherwise.
Di-electronic rates were obtained from \cite{gu03b}, supplemented by
those from \cite{arnaudray} (M-shell Fe ions), \cite{shull82} (M-shell
Ni ions) and \cite{arnaudroth} (He- and Li-like C, N, O and Ne).
Collisional ionization rate coefficients ($C_{DI}$ and $C_{EA}$) are
obtained from \cite{arnaudray} for Fe and \cite{arnaudroth} for all
other elements.

\subsection{Wind models}

The modification described above mean that the code is now capable of
simulating all the necessary atomic physics to compute X-ray
spectra for wind models and several such calculations will be
presented in the Sections below. 

For the moment, we continue to work with the
simply-parameterized, smooth outflow models exactly as described in
Paper~I. This approach allows us to investigate outflow signatures and
their sensitivity to the major wind parameters in a relatively
simple manner. In the
future, however, we plan to extend our studies
to models that go beyond the smooth,
steady-state flow prescriptions we adopt here. Numerical simulations
clearly suggest that outflows are likely to have complex structure
that will affect the observed spectra (see
e.g. \citealt{proga04,schurch09}) and time variation in the outflow
properties may well have a role in explaining the observed time
dependence of some spectral features.

\section{Test of ionization balance}
\label{sect:iontest}

The most significant improvements in the physics of the current code
compared to that described in Paper~I are the extension to lower
ionization states and the substantially more sophisticated treatment
of the ionization balance. To test these improvements, we have made a
simple comparison with the ionization balance in the well-known
photoionization 
radiative transfer code {\sc cloudy} (version 07.02, last described 
by \citealt{ferland98}). For the test, we used a wind model with very
low column density such that it would be optically thin and every cell
would be illuminated with the pure power-law continuum from the X-ray
point source. A primary photon power-law index of $\Gamma = 1.2$ was
used for the test and the kinetic temperature was forced to be 
$10^6$~K everywhere. Ionization fractions were computed using
the full machinery of our code and the results for Fe ions are plotted in
Figure~\ref{fig:ioncompare} versus an ionization parameter defined as

\begin{equation}
\xi = \frac{L_{0.1 - 50}}{r^2 n_H}
\end{equation}
where $L_{0.1 - 50}$ is the source luminosity between 0.1 and
50~keV. The Fe ionization fractions obtained with {\sc cloudy} for a
low column-density spherical shell are over-plotted for comparison.

\begin{figure}
\epsfig{file=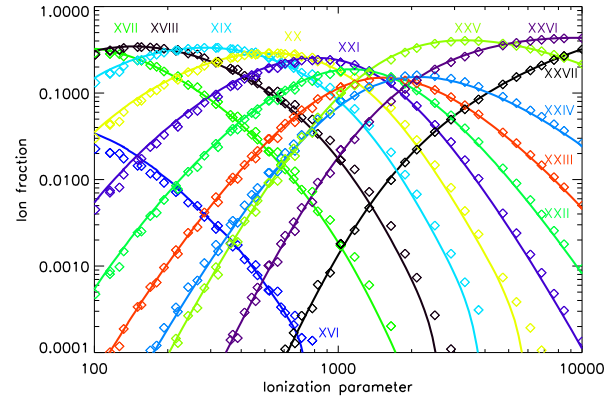, width=8cm}
\caption{
Ionization fractions as a function of ionization parameter 
for Fe {\sc xvi} -- {\sc xxvii} computed for
optically thin conditions with {\sc cloudy} (solid lines) and our
Monte Carlo code (diamond symbols). In both calculations, a power-law
continuum with photon index $\Gamma = 1.2$ and a kinetic temperature
of $T_{e} = 10^6$~K were adopted.
}
\label{fig:ioncompare}
\end{figure}

The agreement between the ionization calculations in the two codes is
very good, better than 10 per cent for most of the K- and L-shell
ions of iron. This precision is comparable to the accuracy of the
source atomic data. Our calculations become less reliable for
ionization parameters much below $\xi \sim 100$~ergs~cm~s$^{-1}$ since our
current atomic data set does not contain all the necessary ions (in
our test calculation, Fe~{\sc ix} -- which we neglect --
has a significant role in the {\sc cloudy} ionization balance below $\xi \sim 50$~ergs~cm~s$^{-1}$).

\section{Example calculations}
\label{sect_example}

\subsection{Model parameters}

Here we present details of two outflow models computed with the
improved code. The first, Model~A has the same outflow parameters as
the example model described in Paper~I. However, the physical
properties of the flow differ slightly from Paper~I since the kinetic
temperature is no longer a parameter but is determined as described in
Section~{\ref{sect:get_tb}. Also,  since the new code version
allows for a finite spatial extent of the primary X-ray
emission region and occultation of the
scattered/reprocessed light (see Section~\ref{sect:disk}), the relative
amplitudes of the direct and scattered/reprocessed components of the
spectrum are altered (see Section~\ref{sect:spec}).

Our second model, Model~B, has parameters chosen to illustrate the
effects of lower ionization state material in the wind as can now be
treated with the improved code version. We adopted a black-hole
mass of $10^7$~M$_{\odot}$ 
and an outflow that is wider than those considered
in Paper I, extending across a factor of three range of launching
radius on the disk (50 -- 150 $r_{g}$). Since the ionizing source is
centrally concentrated this means that the ionization gradient across
the wind (which was already present in Paper I) now extends to even
lower ionization states around the outer edge. We have also chosen a
more gradually accelerated outflow with lower terminal
velocity. Specifically, we adopt a larger value of the 
velocity-law acceleration length parameter $R_{v}$ (which specifies how far downstream the
outflow reaches one half of the terminal speed; see equation 1 of
Paper~I) and a smaller value of $f_{v}$
(which relates the terminal speed to the escape speed at the base of a
streamline; see equation 2 of Paper~I).
The X-ray source parameters and the wind mass-loss rate for
Model~B were
motivated by previous studies of the bright quasar PG1211+143 with
which more detailed comparisons are made in Section~\ref{sect_pg12}.
The primary X-ray source luminosity and the total bolometric
luminosity were chosen to give 
ratios to the Eddington luminosity ($L_{\mbox{\scriptsize Edd}}$) similar to those
inferred for PG1211+143: adopting a typical 2 -- 10~keV luminosity
of $10^{44}$~ergs~s$^{-1}$ and bolometric luminosity of
$4 \times 10^{45}$~ergs~s$^{-1}$
 \citep{pounds03} yields $L_{X}/L_{\mbox{\scriptsize Edd}} \sim
0.02$ and $L_{\mbox{\scriptsize bol}}/L_{\mbox{\scriptsize Edd}} \sim
0.8$ for a black hole mass $\sim 4 \times 10^7$~M$_{\odot}$
\citep{kaspi00} and accretion efficiency of $\sim 0.06$.
The power-law index ($\Gamma \sim 1.8$) 
and mass-loss rate ($\dot{M} \sim \dot{M}_{\mbox{\scriptsize Edd}}$)
were also chosen to be roughly appropriate for PG1211+143, as
motivated by the discussion of \citet{pounds03}.

The complete set of model parameters is given in
Table~\ref{tab_param}; definitions of the geometry and velocity-law
parameters are given in Paper~I.
For both models, we placed the innermost edge for the accretion disk
at the last stable
orbit for a Schwarzschild black hole ($6 r_{g}$) and assumed a
comparable size for the region of X-ray emission (again $6
r_{g}$).

\begin{table*}
\caption{Inputs parameters for the example model.}
\label{tab_param}
\begin{tabular}{lll}\\ \hline
Parameter & Model A & Model B \\ \hline \hline 
mass of central object, $M_{bh}$ & $4.3 \times 10^6$ M$_{\odot}$ &$10^7$ M$_{\odot}$ \\
bolometric source luminosity, $L_{\mbox{\scriptsize bol}}$ & $2.5 \times 10^{44}$
ergs~s$^{-1}$ ($\sim 0.6
L_{\mbox{\scriptsize Edd}}$)& $10^{45}$ ergs~s$^{-1}$ ($\sim 0.8
L_{\mbox{\scriptsize Edd}}$)\\
source luminosity (2 -- 10 keV), $L_{X}$ & $10^{43}$
ergs~s$^{-1}$ ($\sim 0.02
L_{\mbox{\scriptsize Edd}}$) & $2.5 \times 10^{43}$
ergs~s$^{-1}$ ($\sim 0.02
L_{\mbox{\scriptsize Edd}}$)\\
source power-law photon index, $\Gamma$ & $2.38$ & $1.8$ \\
range of source photon energies in simulation & 0.1 -- 511~keV & 0.1 -- 511~keV \\
size of emission region, $r_{er}$ & $6 r_g = 3.8 \times 10^{12}$ cm & $6 r_g = 8.9 \times 10^{12}$ cm \\
inner radius of disk,$r_{d}$ & $6 r_g = 3.8 \times 10^{12}$ cm &  $6 r_g = 8.9 \times 10^{12}$ cm \\
inner launch radius, $r_{\mbox{\scriptsize min}}$ & 100 r$_g = 6.4
\times 10^{13}$ cm & 50 r$_g = 7.4 \times 10^{13}$ cm \\
outer launch radius, $r_{\mbox{\scriptsize max}}$ & $1.5
r_{\mbox{\scriptsize min}}$ & $3 r_{\mbox{\scriptsize min}}$ \\
distance to wind focus, $d$ & $r_{\mbox{\scriptsize min}}$ & $r_{\mbox{\scriptsize min}}$ \\
terminal velocity parameter, $f_{v}$ & 1.0 & 0.5 \\
velocity scale length, $R_{v}$& $r_{\mbox{\scriptsize min}}$ & $3 r_{\mbox{\scriptsize min}}$\\
velocity exponent, $\beta$& 1.0 & 1.0 \\
launch velocity, $v_{0}$ & 0.0 & 0.0 \\
wind mass-loss rate, $\dot{M}$ & 0.1 M$_{\odot}$
yr$^{-1}$ ($\sim 0.6 \dot{M}_{\mbox{\scriptsize Edd}}$)$^{a}$
& 0.38 M$_{\odot}$
yr$^{-1}$ ($\sim \dot{M}_{\mbox{\scriptsize Edd}}$)$^{a}$\\
mass-loss exponent, $k$ & -1.0 & -1.0 \\ 
outer radius of simulation grid & {$5 \times 10^{16}$ cm}& {$5 \times 10^{16}$ cm}\\
3D Cartesian RT grid cells & {$180 \times 180
  \times 180$} & {$180 \times 180  \times 180$}\\
2D wind grid zones & {$100 \times 100$} &{$100 \times 100$} \\
\hline
\end{tabular} \\
$^{a}$ The Eddington accretion rate, $\dot{M}_{\mbox{\scriptsize
    Edd}}$ was computed assuming a radiative efficiency of six percent.
\end{table*}

\subsection{Thermal and ionization structure}

Figure~\ref{fig:thermalstate} shows the distribution of kinetic
temperature and Fe ionization states in the example models.
For Model~A,
the spatial variation of the ionization state is qualitatively similar
to that discussed in Paper~I: the inner edge of the wind
is almost fully ionized and gradients of decreasing ionization occur
both along the outflow and across the base of the flow. Model~B is similar but,
owing to the
larger radial extent of the flow launching region in this model, the
gradient across the flow is particularly well-developed such that a
substantial region on the outside of the wind is dominated by L-shell
Fe ions.
As expected, the regions of the wind that are most strongly ionized
are also most strongly heated by the X-ray source leading to
significant kinetic temperature variations across the flow (right
panels of Figure~\ref{fig:thermalstate}).

\begin{figure*}
\epsfig{file=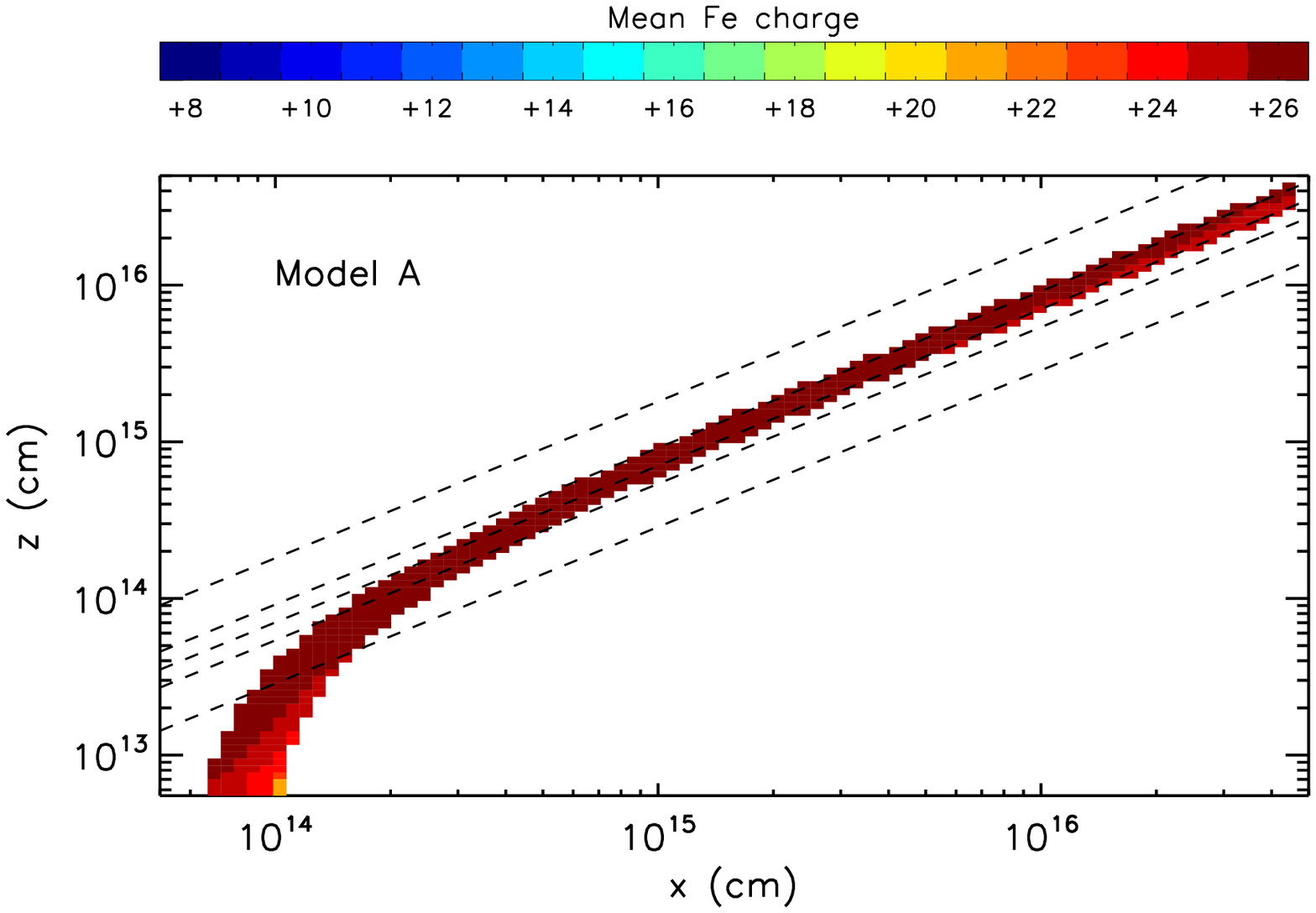, width=5.8cm,angle=90}
\hspace{-1.1cm}
\epsfig{file=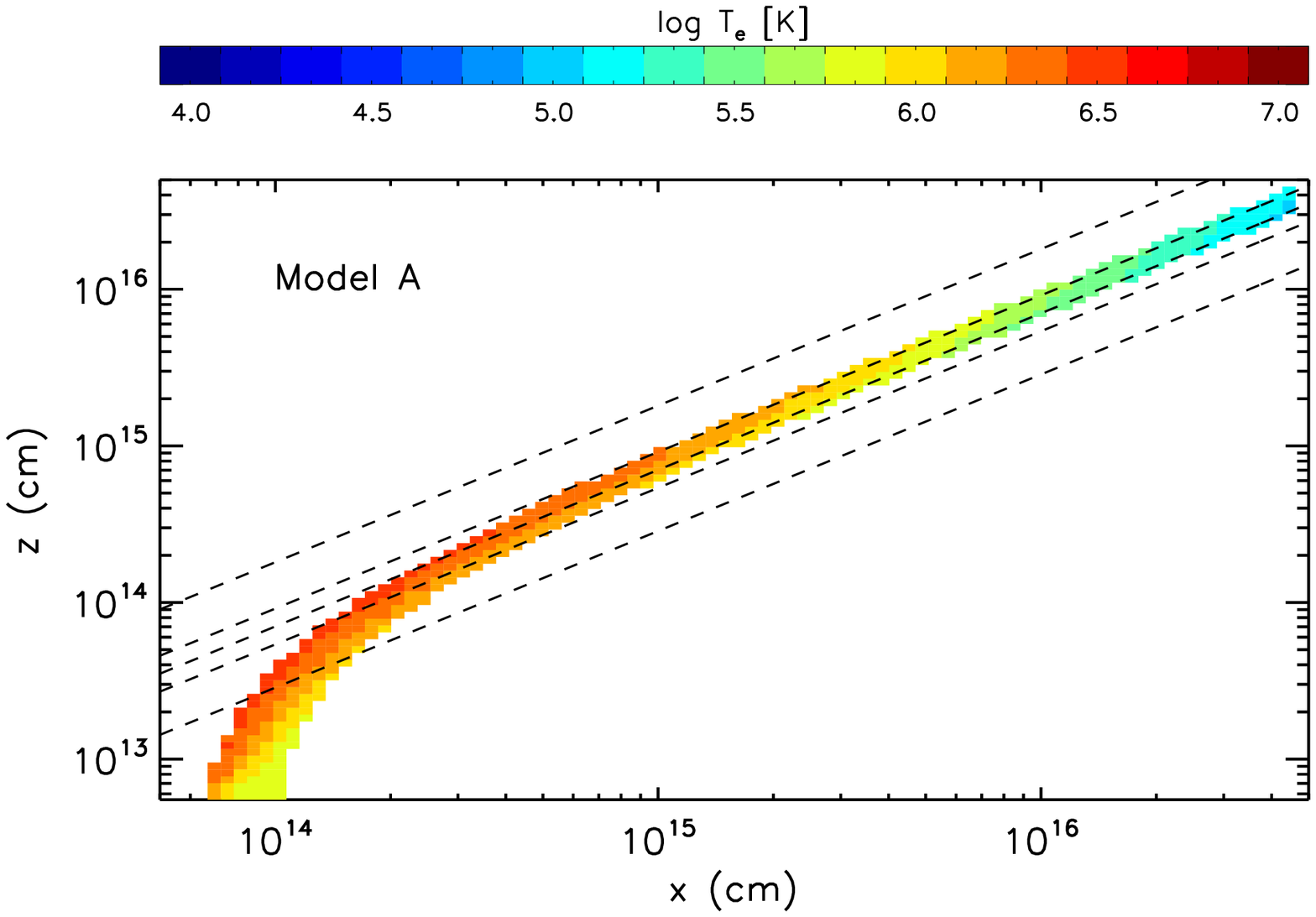, width=5.8cm,angle=90}\\
\epsfig{file=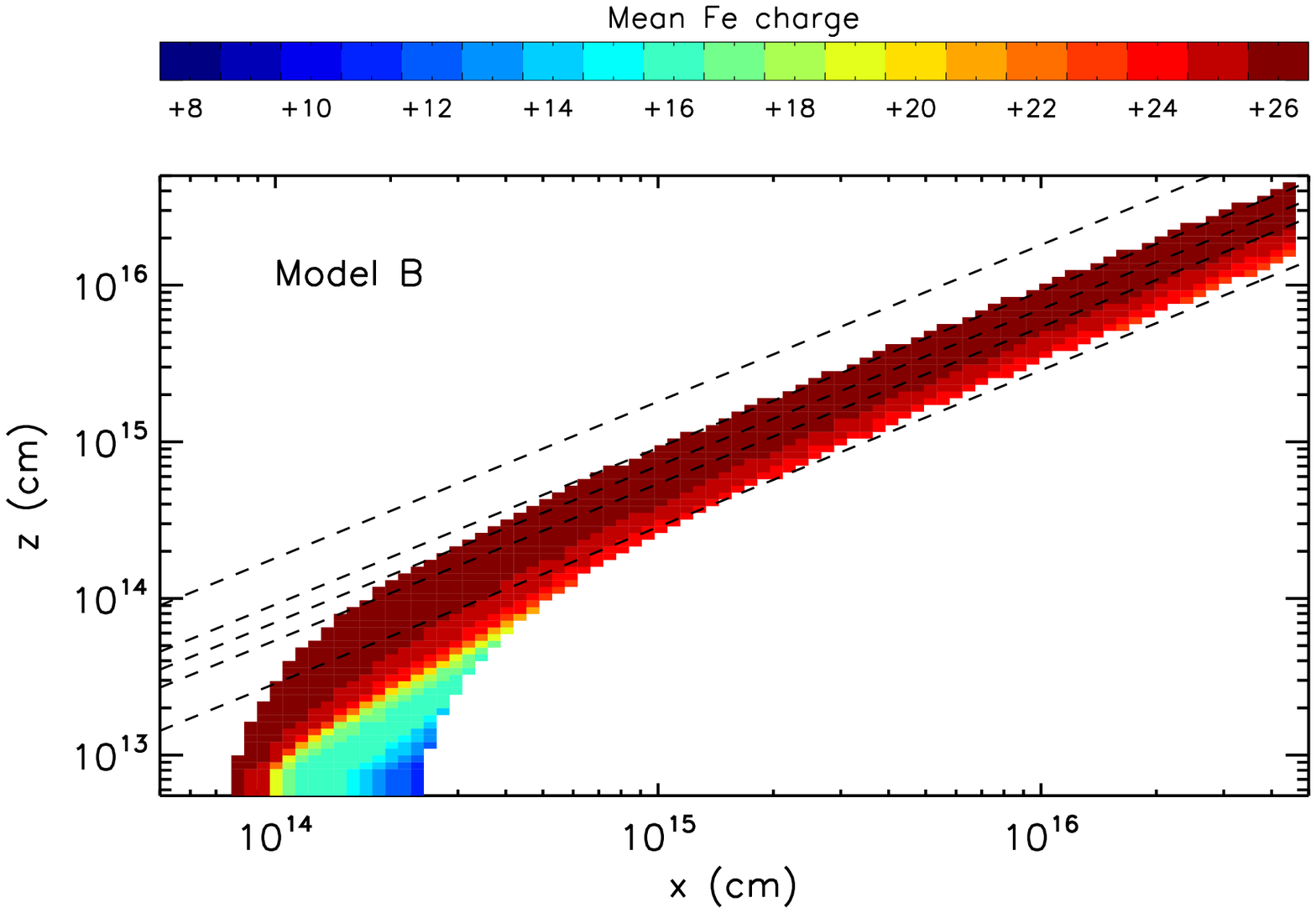, width=5.8cm,angle=90}
\hspace{-1.1cm}
\epsfig{file=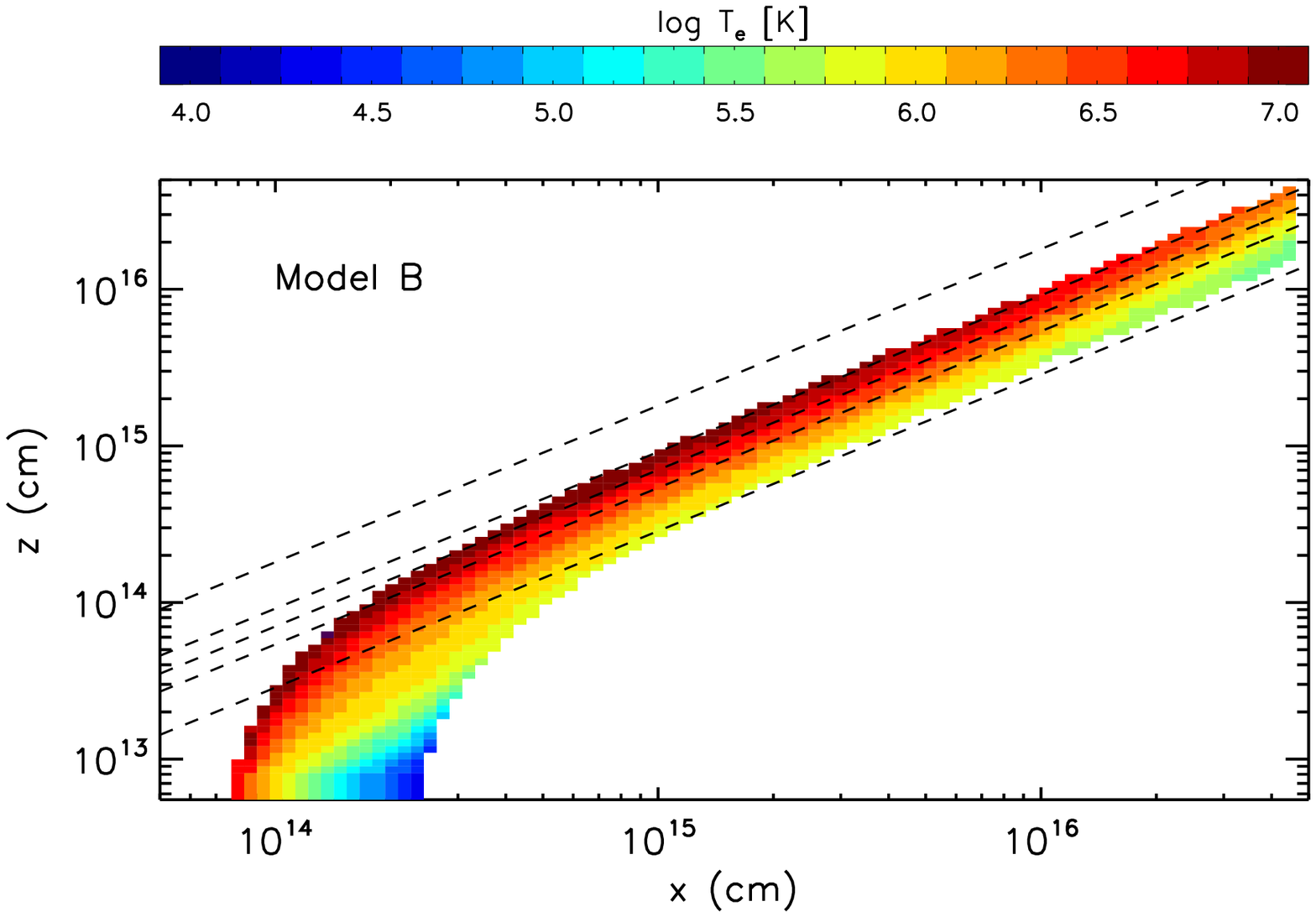, width=5.8cm,angle=90}
\caption{
Distribution of mean Fe ionization state (left) and 
kinetic temperature (right) for
Models~A (top) and B (bottom).
The dashed lines indicate the five
lines-of-sight for which spectra are shown in Figures~3 and 4.
}
\label{fig:thermalstate}
\end{figure*}

\subsection{Computed spectra}
\label{sect:spec}

For both the example models, we computed emergent spectra for twenty
lines-of-sight uniformly sampling the orientation ($0 < \mu <
1$, where $\theta = \cos^{-1} \mu$ is the 
angle between the line-of-sight and the
rotation axis). Figures~\ref{fig:examplespec_old} and
\ref{fig:examplespec} show the computed spectra for five of these 
lines-of-sight for Models~A and B, respectively. 
The orientations of the chosen lines-of-sight are indicated by the dashed
lines in Figure~\ref{fig:thermalstate}.
As in Paper~I, the propagation
histories of the Monte Carlo quanta are used to divide the emergent
spectrum into a direct component (representing photons that reached
the observer without any interactions) and a scattered/reprocessed
component derived from the quanta that underwent at least one
interaction in the wind. These contribution are separately plotted in
the left panels of Figures~\ref{fig:examplespec_old} and \ref{fig:examplespec}.
The right panels show the 2 -- 10 keV region of the spectra in
greater detail. 

\begin{figure*}
\epsfig{file=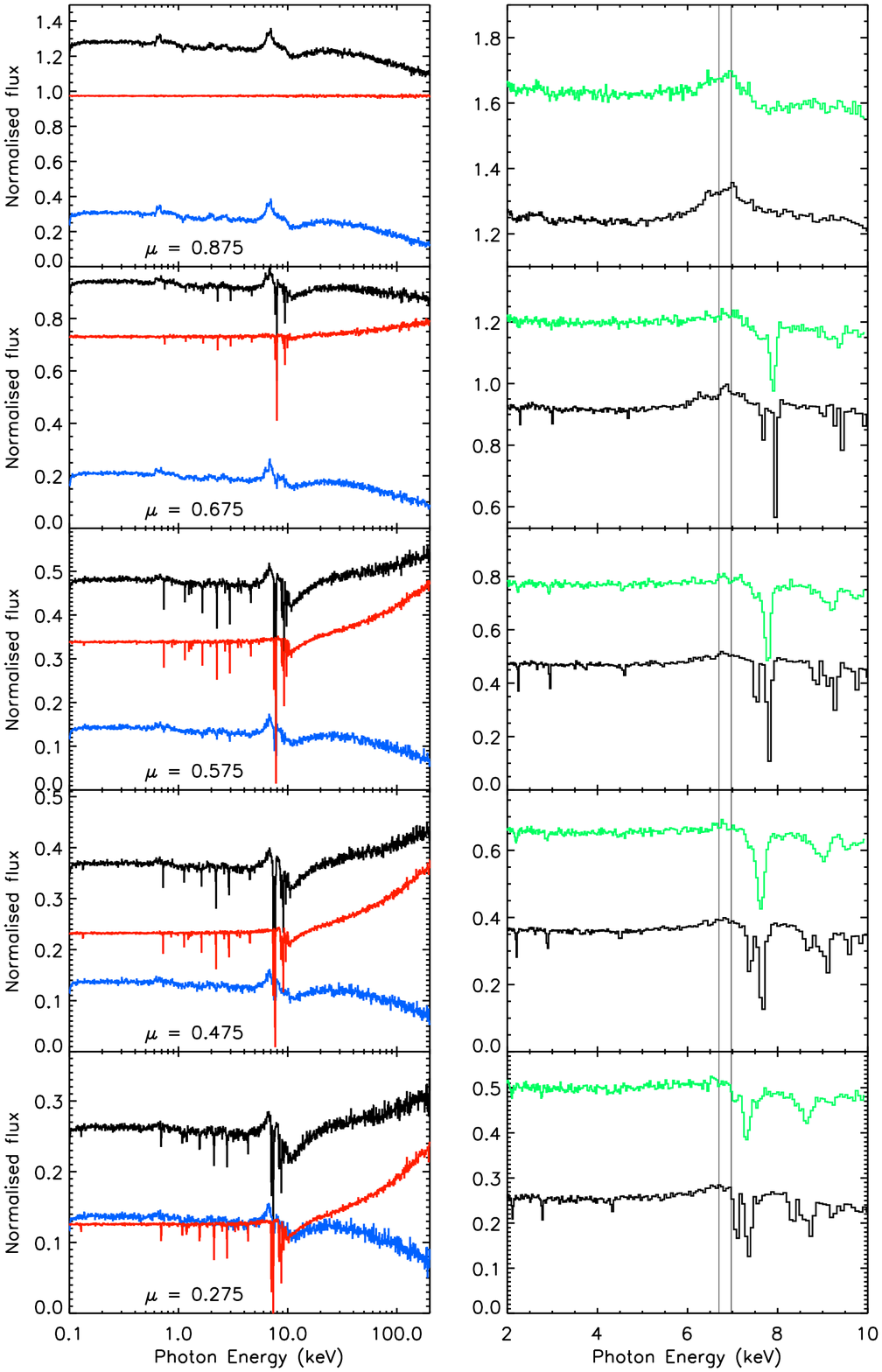, width=12cm}
\caption{
Spectra for five observer lines-of-sight computed from Model~A 
(from top to bottom, $\mu = 0.875$, 0.675, 0.575, 0.475 and 0.275
where $\theta = \cos^{-1} \mu$ is the angle of the line-of-sight relative to the polar
axis). The left panels show the spectrum from 0.1 -- 200 keV while the
right panels show the 2 -- 10 keV region in detail for the same lines-of-sight.
Note that the abscissa is plotted logarithmically in the left panels but linearly in
the right panels.
In the left panels, the total spectrum is shown in black, the spectrum
of direct photons in red and the scattered/reprocessed spectrum in
blue. 
In the right panels, the total spectrum computed with the new code
version is plotted in black while the spectra from the example model in
Paper~I are shown in green, for comparison.
The vertical lines indicate the mean rest energies
of the Fe~{\sc xxv}/{\sc xxvi} K$\alpha$ transitions ($\sim$6.7/6.97~keV).
All the spectra are normalised to the incident X-ray spectrum which
has a photon power-law index, $\Gamma$=2.38. Monte Carlo noise from
the simulations is
present in all the spectra and is responsible for the small-scale
fluctuations.
}
\label{fig:examplespec_old}
\end{figure*}

\begin{figure*}
\epsfig{file=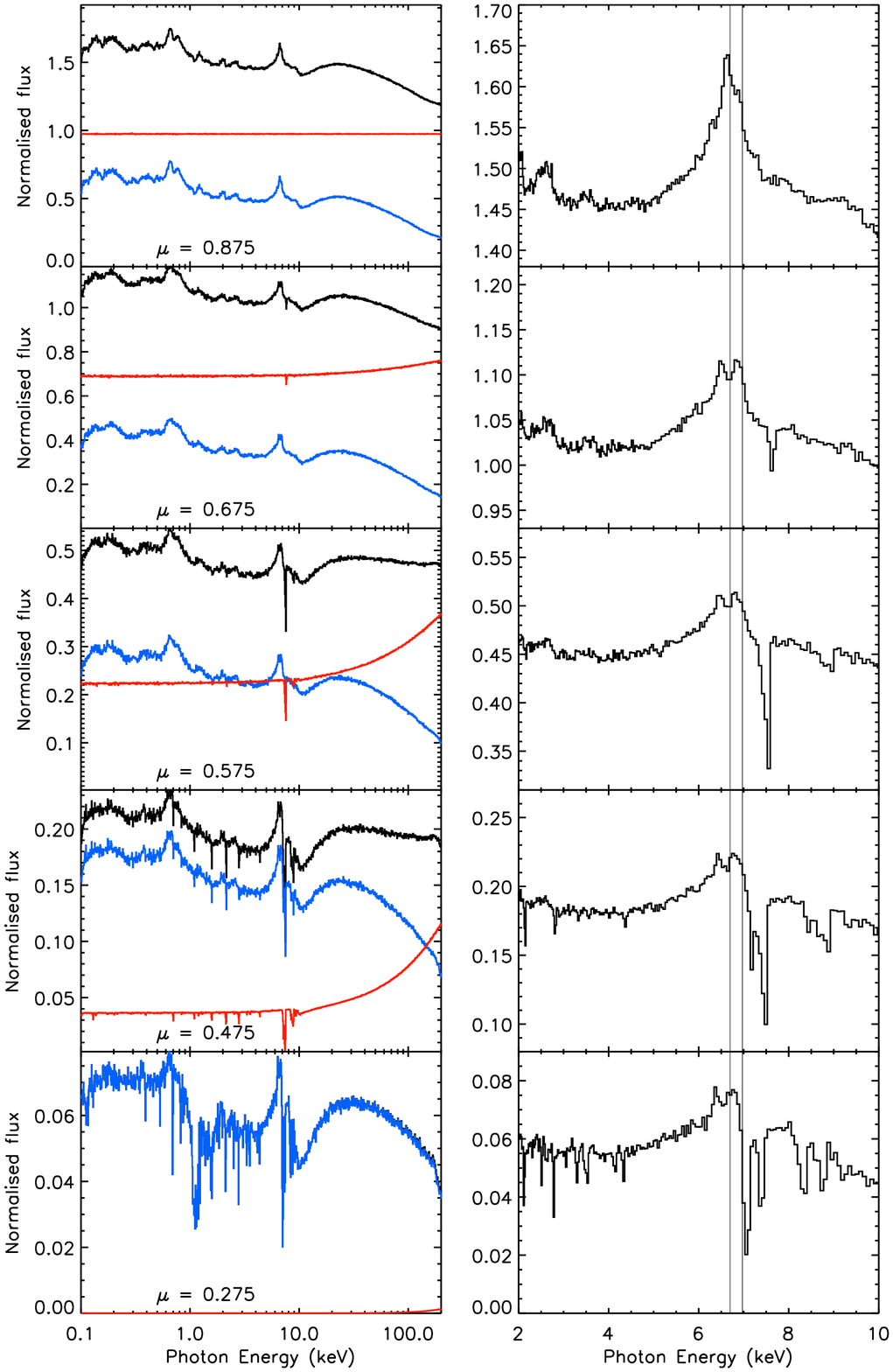, width=12cm}
\caption{
As Figure 3 but showing spectra from Model~B. No comparison spectra
are shown in the right panels.
All the spectra are normalised to the incident X-ray spectrum which
has a photon power-law index, $\Gamma$=1.8.
}
\label{fig:examplespec}
\end{figure*}

The first line of sight shown for both models (top panels of
Figures~\ref{fig:examplespec_old} and \ref{fig:examplespec}) 
corresponds to an observer
at sufficiently low inclination that no portion of the wind obscures
the primary X-ray source. 
Thus the direct component of radiation is
unaffected by the wind (red line in top left panels). However, since
the outflow scatters light from other lines-of-sight and produces its
own thermal emission, the direct light is supplemented by a
significant component of scattered/reprocessed radiation. As one would
expect, this scattered/reprocessed spectrum is qualitatively similar
to that obtained from standard disk reflection models
(e.g. \citealt{ross05}). 
In particular, a moderately strong Fe~K$\alpha$ emission 
line appears in both models. 
This is predominantly formed by Fe~{\sc xxv} and {\sc
  xxvi} since most of
the wind reflection seen from this line-of-sight occurs in the
highly-ionized inner surface of the outflow. The emission line profile
is Doppler broadened and, owing to the substantial outflow velocities, 
develops an electron scattering wing \citep{auer72,laurent07} that
skews the profile to the red (for both models, 
it extends down to $\sim 5$~keV for this line-of-sight).
At softer energies, the spectrum of scattered Monte Carlo quanta also
contains a variety of other emission features. These are considerably
stronger in Model~B owing to the greater range of ionization
conditions in the wind. In this model, weak but distinct
K$\alpha$ emission lines of O, Si and S 
($\sim 0.65, 2.0$ and 2.6~keV, respectively) are present together with
forests of blended lines from the L-shell ions of Fe.
The emission profiles of these features are broadened by the same
mechanisms which affect the Fe K line leading to rather complex
spectra. 
The scattered/reprocessed spectrum also adds a
distinct ``Compton hump'' to the total spectrum, causing it to have a
broad peak around 20 -- 30 keV; at even harder energies the scattered/reprocessed
spectrum bends down (owing to the energy dependence of the Compton
cross-section) such that the flat, primary X-ray spectrum becomes
increasingly dominant above about 50~keV.
For this inclination angle, the spectrum computed for Model~A is very
similar to that obtained for the same outflow parameters using the
code version in Paper~I (see upper right panel of
Figure~\ref{fig:examplespec_old}). The main difference is a modest
offset in normalization that arises from the reduced amplitude of the
scattered component of radiation owing to occultation by the accretion
disk which is included in the new calculations (see Section~\ref{sect:disk}).

The second line of sight considered ($\mu = 0.675$)
passes through the upper-most
layers of the outflow models which are hottest and most
highly-ionized. The remaining three lines-of-sight ($\mu = 0.575$, 0.475
and 0.275) pass through
increasingly denser and cooler parts of the wind such that the spectra
become more complex.
As in Paper~I, lines of sight through the wind show
weak, narrow, blueshifted absorption features associated with highly ionized
material, most importantly the K$\alpha$ lines of Fe~{\sc xxv} and
{\sc xxvi}. In Model~A, the spectra for $\mu = 0.675$ (and lower) 
are characteristically similar to
that of the example model in Paper~I (see comparison in
Figure~\ref{fig:examplespec_old}) but there are some differences in
detail. In particular, with the improved code version 
the amplitude of the component of scattered/reprocessed
light is reduced by disk occultation, the mean ionization state is
slightly lower and the absorption features are sharper. This last
effect arises because of the angular binning of MC quanta used to
extract the spectra in Paper~I: since the line-shifts depend on
inclination, spectra obtained via angular binning will tend to have
slightly smeared line profiles.

In Model B,
the 
$\mu = 0.675$
line of sight  
has an
integrated hydrogen column density of $N_{H} \sim 2 \times 10^{23}$~cm$^{-2}$, only a
factor of a few smaller than that 
suggested for the most highly ionized material in the spectrum of
PG1211+143 ($\sim 5 \times 10^{23}$~cm$^{-2}$, \citealt{pounds03}). 
However, the material
along this line of sight is hot and almost fully ionized 
(see Figure~\ref{fig:thermalstate}) such that Compton scattering is the
only important opacity source. This reduces the flux of the direct
component of the radiation and causes it to curve upwards at high
energies, a consequence of the photon-energy dependence of the Compton
cross section. Although the scattered/reprocessed component of the
light still shows a significant Compton hump, this upwards curvature of
the direct component partially cancels it out leaving only a modest
bump in the total spectrum. The only significant absorption
feature imprinted by the wind for this line of sight is from the trace
population of Fe~{\sc xxvi} and only the K$\alpha$ line is
strong enough to be plausibly detectable with current instruments. The
emission features in the scattered/reprocessed spectrum are
qualitatively very similar to those for the lower inclination angle
described above except that the features are slightly broader thanks
to the greater line-of-sight components of both the outflow and
rotational velocity fields for this orientation. In addition, the
Fe~K$\alpha$ line shows two distinct peaks, one close to the rest
K$\alpha$ energy of
Fe~{\sc xxvi} and the other to that of Fe~{\sc xxv} and the high L-shell ions. 

The third to fifth lines-of-sight for which spectra are 
shown for Model B (Figure~\ref{fig:examplespec})
have sufficiently high column densities
($N_{H} \sim 10^{24}$, $\sim 4 \times 10^{24}$ and 
$\sim 10^{25}$~cm$^{-2}$, respectively) that the direct
component of radiation is almost completely blocked except at rather
high energies ($\simgt 30$~keV). For the highest inclination angle
considered ($\mu = 0.275$), the reprocessed component of light
dominates even well above 100~keV such that the total spectrum of the
wind shows a
downturn at these energies, a property that can be constrained by
high-energy observations (see discussion by \citealt{fabian95}). 
However, we stress that for the lower inclination angles in this model,
and indeed all the angles considered for Model~A, 
the direct component rises sufficiently rapidly with energy
that there is still significant flux emerging above 200~keV.

For the $\mu = 0.575$, 0.475 and 0.275 lines-of-sight, the Model~B spectrum is progressively
more strongly affected by absorption in the wind. The Fe~K$\alpha$
absorption line becomes deeper and gradually switches from being
dominated by Fe~{\sc xxvi} to Fe~{\sc xxv} at higher inclination
angles. The associated K$\alpha$ emission line remains strong and its
apparent red wing becomes increasingly extended for high inclination angles.
As expected, the Fe~K$\alpha$ absorption is also accompanied by blueshifted
absorption by Fe~K$\beta$ ($\sim 9.0$~keV) and Ni~K$\alpha$ (Ni~{\sc
  xxvii} at $\sim 8.5$~keV and {\sc xxviii} at $\sim 8.8$~keV). There is
also significant absorption around the blueshifted Fe~K edge which,
although somewhat smeared out by the Doppler shifts, causes the
distinct drop in flux at around 10~keV.

At softer energies, blueshifted absorption features associated with
the abundant light elements become increasingly prominent as the
inclination angle is increased. 
For an inclination of $\mu = 0.475$ (fourth row in
Figure~{\ref{fig:thermalstate}), weak narrow K$\alpha$ absorption
  lines of 
S~{\sc xvi} ($\sim 2.8$~keV), Si~{\sc xiv} ($\sim 2.2$~keV), Mg~{\sc
  xii} ($\sim 1.6$~keV), Ne~{\sc x} ($\sim 1.1$~keV) and O~{\sc viii} ($\sim 0.71$~keV)
are all present but their equivalent widths are small, generally
less than a few eV. Weak Fe~{\sc xxiv}
lines also appear at $\sim 1.2$ and 1.3~keV due to blueshifted
transitions between its ground 2p$^5$ configuration
and states of the excited 2p$^{4}$~($^{3}$P)3s configuration.
At the higher inclination angle of $\mu = 0.275$ (fifth row in 
Figure~\ref{fig:examplespec}), the absorption spectrum is even
more complex. The K$\alpha$ transitions mentioned above are all still
present and generally somewhat stronger. Very weak K$\alpha$ lines due
to Ar and Ca also manifest ($\sim 3.6$ and 4.4~keV, respectively), as
do the K$\beta$ lines of the H-like ions of Mg, Si and S. Although
O~{\sc viii} K$\beta$ absorption is also significant in the
simulation, this feature is blended with comparably strong lines of
Fe~{\sc xviii} and {\sc xix}. Deep absorption occurs between 1.0 and
1.3~keV owing to the forest of lines arising in transitions from the low-lying $n=2$ configurations of Fe and Ni L-shell ions.
Furthermore, there is also significant bound-free absorption by the
ground states of the K-shell ions of O, Mg, Si and S and of the
L-shell ions of Fe that causes a drop in the flux above $\sim 1.5$~keV.

\section{Preliminary comparison with PG1211+143}
\label{sect_pg12}

Thanks to the extension of the capabilities of the code to deal with
lower ionization conditions, we can now use the models to compare
much more realistic theoretical spectra with observations to
establishing how readily outflow models can explain observed
spectroscopic features and, if real, what the physical properties of
these flows might be. For a fully detailed comparison a large grid of
theoretical models covering the physically interesting parameter space
must be computed. This will be an important next step in our studies
but we begin here by making a preliminary comparison with the observed
spectrum of the bright quasar PG1211+143.

\subsection{Observational data}

Amongst the first direct evidence for highly ionized massive outflows
from AGN was a report by \cite{pounds03,pounds05} based on the
analysis of 
{\it XMM-Newton}
X-ray spectra of PG1211+143. They reported the detection of several
strong absorption features which they associated with blueshifted
K$\alpha$ transitions of C, N, O, Ne, Mg, S and Fe. To quantify the absorption
features in the (1 -- 10~keV) {European Photon Imaging Camera} (EPIC) pn spectrum they fit an {\sc xstar} \citep{kallman01}
absorption model and found that a high column density ($N_{H}
\sim 5 \times 10^{23}$~cm$^{-2}$) and a high ionization parameter
($\log (\xi / \mbox{ergs~cm~s$^{-1}$}) \sim 3.4$) were required to explain both the claimed Fe and S~K$\alpha$
absorption lines as originating in a rapid ($v \sim 0.08$c)
outflow. Based on the proposed identification of the lower ionization
Mg line in the EPIC pn spectrum and lines attributed to the lighter
elements in their contemporaneous {Reflection Grating
  Spectrometer} (RGS) spectra, they also postulated
that additional absorbing components must be present in the outflow
with significantly lower column density and ionization parameter
(although comparable outflow speeds). 
Broad
emission was also identified in the EPIC pn spectrum, extending from 
$\sim 3$ to 7~keV. 
Although this emission feature can be 
fit as 
an extreme relativistic Fe~K$\alpha$ emission line
originating from reflection by the
inner most regions of an ionized accretion disk \citep{pounds03},
it has more recently been suggested that it 
may be P~Cygni emission physically associated with 
the blue-shifted K$\alpha$ absorption (e.g. \citealt{pounds09}) as
would be expected from a wind with a large covering fraction. 

Subsequent re-observations of PG1211+143 with {\it XMM-Newton} were
made in 2004 \citep{pounds07} and 2007 \citep{pounds09}. These showed
that, although the spectral properties of PG1211+143 are significantly
time-variable, the fast outflow of highly ionized gas is persistent
and confirmed that it is likely to be an energetically important
component of the system.

Given the complexity of the available data for PG1211+143 and the wide
range of spectroscopic features identified in the 2001 {\it XMM} data first reported
by \cite{pounds03}, this object provides an excellent point of
comparison with our synthetic spectra for outflows. In particular,
while previous efforts to quantify the spectra have generally adopted
distinct fit components for the claimed absorption lines and the
apparently broad emission redward of $\sim$7~keV, our wind models allow us
to explore the relationship between these features with a set of
physical models.

The X-ray spectrum of PG1211+143 shows a strong excess for energies
$\simlt 1$~keV relative to an extrapolation of a power-law fit to the
harder energy spectrum (see e.g. \citealt{pounds03}).
The origin of soft excesses in
AGN spectra remains unclear: although it has been suggested that they
may be a consequence of absorption by relatively low ionization
material and data showing strong soft excesses 
(including those of PG1211+143; see \citealt{pounds07,pounds09})
have been successfully fit assuming an absorption-dominated origin,
the geometry and origin of the absorbing material is not
well-constrained. 
In view of this
uncertainty, we restrict our fitting to the 2 -- 10~keV spectrum of
PG1211+143 to avoid
biasing our fit by a choice of particular model for the origin of the 
soft excess. However, we will qualitatively discuss the lower-energy absorption
features predicted by the model compared to those 
reported in the PG1211+143 spectra.

In the following, we will make quantitative comparisons with the
data set compiled by \cite{pounds09} by stacking {\it XMM} EPIC pn
data from 2001, 2004 and 2007. We have rebinned the data to the
energy-dependent half-width at half-maximum (HWHM) of the EPIC pn 
instrument.

\subsection{Choice of model parameters}

We began by making a comparison of the stacked 2 -- 10~keV data of
PG1211+143 with our example spectra (described in
Section~\ref{sect_example}). As discussed above, Model~B already
indicated that many of the discreet features identified in the 
spectrum of PG1211+143 can qualitatively be reproduced in our theoretical
spectra. To produce a strong, significantly blueshifted Fe
{\sc xxv}/{\sc xxvi} absorption feature, the example models suggest
that a line-of-sight passing through some portion of the inner, highly
ionized edge of the flow is required ($\mu \sim 0.4$ --
0.6 for the example model geometric parameters). Although Model~B 
already produces strong K$\alpha$ absorption, the blueshift is
slightly too small compared to that suggested for PG1211+143 
\citep{pounds03,pounds05,pounds06}. Therefore, we first increased the outflow
velocity parameter ($f_{v}$) from $1/2$ to $2/3$, a value intermediate
between those used for the two example models.

Upon quantitative comparison, the Model~B spectra also differ in detail
from the observations of PG1211+143 in that the lower ionization
features are somewhat too weak for the appropriate inclination angles
(e.g. the equivalent width of
the S~{\sc xvi} Ly$\alpha$ line is only $\simlt 5$eV in the model 
while \cite{pounds05} and \cite{pounds06} report equivalent widths of
$\sim 40$ and $\sim 24$eV, respectively).
Thus, models with lower typical
ionization state are desirable. 
In principle, a full grid of models exploring all the outflow
parameters should be explored to find the best match to the observed
spectrum. Here, however, we restrict ourselves to varying parameters
that are expected to directly affect the mean ionization state.
We therefore
created a small grid of six models that explores the effects of
higher wind mass-loss rates ($\dot{M}$) and longer velocity-law scale
lengths ($R_{v}$) while keeping all other parameters the same as for Model~B. 
Relatively modest numbers of Monte Carlo quanta were
used for these simulations in order to reduce the computation demands
(only $\sim 10^6$ packets were used for the calculation of the spectra).

The quantitative effect of $\dot{M}$ on the models was discussed in
Paper~I and that discussion remains applicable here. In particular,
increasing $\dot{M}$ leads to higher densities and therefore lower
ionization states for a fixed X-ray source luminosity. In general, it
also leads to stronger absorption features thanks to the increased
column density for a fixed line of sight. Our grid of models explored
$\dot{M}$-values of 0.38, 0.76 and 1.52~M$_{\odot}$~yr$^{-1}$
corresponding to $\sim 1$, 2 and 4 $\dot{M}_{\mbox{\scriptsize Edd}}$ for our
  adopted value of $M_{\mbox{\scriptsize bh}} \sim 10^7 M_{\odot}$. 

The second parameter varied, $R_{v}$, affects the spectrum for two
distinct reasons. First, a larger value of $R_{v}$ means that the flow
accelerates more gradually such that the material in the accelerating
region is more dense for fixed values of $\dot{M}$ and
$f_{v}$. Secondly, the larger value of $R_{v}$ reduces the outflow
velocity-gradient such that the Sobolev optical depths of the line
transitions become larger in the acceleration region. For our grid we
considered $R_{v}$-values of 3 and 5~$r_{\mbox{\scriptsize min}}$.

As when making our comparison with the average spectrum of Mrk~766 in
Paper~I, we constructed a multiplicative
table (``mtable'') model for use with {\sc xspec} (version 11, \citealt{arnaud96}) 
from the ratio of our computed spectra to the primary power-law
adopted in the radiative transfer simulations. We then attempted to fit the
PG1211+143 data using a model comprising of a power-law (with
free normalization [$N$] and power-law index [$\Gamma$]) combined with this mtable
(which contains a total of 120 spectra from the grid of 3 $\times$ 2
$\times$ 20 in $\dot{M}$, $R_{v}$ and $\mu$). Given the strong but complex
dependence of the spectral features to the line-of-sight ($\mu$) we
did not allow this parameter to be fit directly but stepped through
the $\mu$-values of the grid manually, freezing this parameter and
then fitting the remaining four quantities ($N$, $\Gamma$, $\dot{M}$
and $R_{v}$). 
We did not allow for any component of systematic
error in any of our fits as this is not expected to have a significant effect
in the fit to the 2 -- 10~keV region (see the discussion in \citealt{miller09}).
Although other physical processes are expected to have a
role in shaping the X-ray spectra of AGN (in particular, some degree
of disk reflection and some warm absorber component seem all but
inevitable), we did not include any additional component in this fit
since our objective is to establish whether our disk wind models alone could be
the dominant component that accounts for the major spectra features in the
2 -- 10~keV band.

For this model, the best fit was obtained for $\mu = 0.525$ (frozen),
$\Gamma = 1.9$, $\dot{M} = 0.68$~M$_{\odot}$~yr$^{-1}$  and $R_{v} =
3.2 \;
r_{\mbox{\scriptsize min}}$ and yielded $\chi^2/$d.o.f of
163/132. Although an imperfect fit, this suggests that the majority of
the spectroscopic features are well accounted for by the model (a pure
power-law fit to the same data yields $\chi^2/$d.o.f of 389/134).

To make a detailed comparison, we computed spectra for eighteen more models
adopting parameters close to those favoured from the {\sc xspec}
fits and extending the range in $R_{v}$ 
($\Gamma = [1.7,1.9]$, $R_{v} = [3,5,8]r_{\mbox{\scriptsize min}}$, $\dot{M} = [0.53,0.65,0.76]$~M$_{\odot}$~yr$^{-1}$). For these simulations a larger number of
MC quanta ($\sim 7 \times 10^6$) were used to compute the spectra. The
spectra obtained from each of these simulations are qualitatively
similar although the amplitude of the spectra features varies
slightly. 
We fit spectra from these eighteen models to the 2 -- 10~keV spectrum in
a similar manner to that described above -- the best fitting spectrum
obtained from this model grid ($\chi^2/$d.o.f of 160/132) was for
parameters close to that of the model with
$\Gamma = 1.9$,
$R_{v} = 8r_{\mbox{\scriptsize min}}$ and $\dot{M} = 0.53$~M$_{\odot}$~yr$^{-1}$ (for
$\mu = 0.575$); therefore we use the spectra from this model as the
basis for our comparison (see below).

\subsection{Discussion}

Figure~\ref{fig:pg1211} compares the Fe K$\alpha$ spectral region for
the model with $R_{v} = 8r_{\mbox{\scriptsize min}}$, $\dot{M} =
0.53$~M$_{\odot}$~yr$^{-1}$ 
and $\mu = 0.575$
with the unfolded data of PG1211+143
\citep{pounds09}.
The comparison is shown for the photon energy scale observed at the
Earth, accounting for the source redshift of $z = 0.0809$ \citep{marziani96}.

The agreement between model and data is generally very good and shows 
that the
disk wind paradigm readily produces the important features that
characterize these data -- blueshifted Fe K$\alpha$ absorption and broad
red shifted K$\alpha$
emission with a red-skewed wing -- with strengths comparable to those
observed. In agreement with the data, the model also predicts weak
S~{\sc xvi} absorption around 2.7~keV.
There are, however, some 
clear discrepancies.
For example, 
additional narrow components of emission may
be required in the emission line profile (see discussion by \citealt{pounds09})
 and the model 
seems to predict too little absorption at around 8~keV.
However, none of these details suggest
that the outflow paradigm is inappropriate for these data and may
simply indicate physically distinct contributions to the spectrum. 
E.g. any very narrow emission
components required by the data 
are unlikely to be associated with the high velocity wind
but can be more readily attributed to reflection by low-velocity
material.

\begin{figure}
\epsfig{file=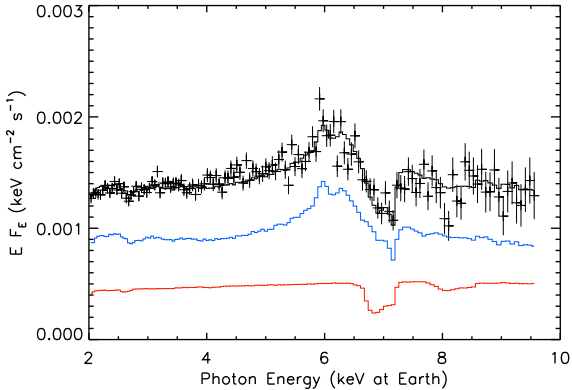, width=8cm}
\caption{
Comparison of the Pounds et al. (2009)
stacked data for PG1211+143 (black crosses) and the wind model (black histogram)
described in the text.
The wind parameters are all identical to Model~B (Table 2) 
except for $\Gamma =  1.9$, $R_{v} = 8
r_{\mbox{\scriptsize min}}$ and $\dot{M} =$ 0.53~M$_{\odot}$~yr$^{-1}$. The
model spectrum shown is for a viewing inclination of $\mu = 0.575$.
The red and blue histograms show the contributions to the model
spectrum from direct and scattered/reprocessed radiation, respectively.
The photon energy scale of the model spectrum has been adjusted to
account for the redshift of PG1211+143 ($z = 0.0809$).
}
\label{fig:pg1211}
\end{figure}

Since the source is clearly complex, fitting the 2 -- 10~keV spectrum 
to a wind model alone yields parameters that
should only by regarded as indicative of those appropriate for PG1211+143.
To make a quantitative comparison of our models with the lower
energy spectra for PG1211+143 would requires significantly 
more complex fits of the data to be performed to address the strong
soft excess and account for any physically distinct, 
lower-ionization contribution to absorption. However, at minimum, the
highly-ionized wind should be able to imprint discreet blueshifted absorption and
broad emission
features in the softer spectrum consistent with those that have been
claimed from previous analyses of the PG1211+143 RGS spectrum
(e.g. \citealt{pounds03,pounds07}). \footnote{We note, however, that
  the exact combination of wind model and
  observer line-of-sight which fit the
  stacked EPIC pn data of \cite{pounds09} (which combines observations from
  2001, 2004 and 2007) is not expected to be 
  perfectly applicable for
  comparison with the RGS absorption features discussed by 
  \cite{pounds03} since it is known that the
  spectroscopic signatures of outflow are time-variable.}

To investigate this, we show in 
Figure~\ref{fig:pg1211_wide} the 10 -- 35\AA~spectra (c.f. figures~6
-- 9 of \citealt{pounds03}) for the same
wind model used in Figure~\ref{fig:pg1211}.
The spectrum is shown for the line-of-sight ($\mu = 0.575$) 
that best fit the Fe~K region of the stacked data of \cite{pounds09}
and also for two higher inclination angles ($\mu = 0.475$ and
0.375).
The wavelength of blueshifted absorption lines for which
identifications were claimed by \cite{pounds03} are indicated in the
figure. Several of these -- in
particular those corresponding to relatively high ionization states
(e.g. Ne~{\sc x}, O~{\sc viii}, N~{\sc vii}, C~{\sc vi}) --
are clearly present in the model with strength and blueshift depending
sensitively on the observer's inclination.
Fe~{\sc xxiv} imprints a clear additional absorption feature for $\mu =
0.475$ -- although they did not discuss this in detail, we note that 
the fit in figure~9 of \cite{pounds03} does include an absorption
line around 11~\AA.
The model also predicts
significantly broadened O~{\sc viii} Lyman $\alpha$ emission,
qualitatively similar to that discussed by \cite{pounds07,pounds09} in their
more recent analyses of the PG1211+143 RGS data.
Thus our
outflow models are able to simultaneously account for 
observable features in both the Fe~K band and the softer-energy parts
of the spectrum and there are good prospects that more detailed
fitting of the complete set of spectral data will allow more robust
constraints to be placed on the outflow model parameters.

The less ionized species (Ne~{\sc ix} and O~{\sc vii}) are present in
only very small quantities in the model and do not imprint lines in the
spectra shown.
The presence of these lines in the data suggests that a wider
range of ionization states than in the model used here 
is required to explain the complete 2001 RGS spectrum. This may be found
by exploring a greater range of our model parameter-space but may also
require us to go beyond our assumption of a smooth, steady-state flow. A
more realistic model incorporating flow inhomogeneities will
inevitably have a wider range of densities, and therefore
ionization state. Such effects will be investigated in future studies.

\begin{figure}
\epsfig{file=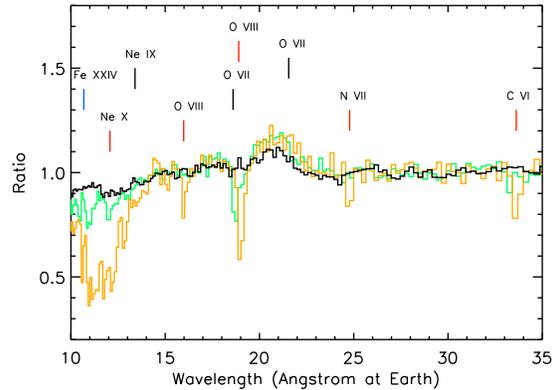, width=8cm}
\caption{
The RGS region of the spectra of the same wind model shown in
Figure~\ref{fig:pg1211}. 
The black histogram shows the model spectrum for $\mu = 0.575$ (same
inclination as used for the fit to the stacked EPIC pn data shown in
Figure~\ref{fig:pg1211}). The green and orange histograms show the spectra
for $\mu = 0.475$ and $0.375$, respectively. Each spectrum is
shown as a ratio to the input pure power-law spectrum which had
$\Gamma = 1.9$ and renormalised to 1.0 around 30\AA.
The identifications with red tick marks
indicate the observed energies of lines identified in the observations of PG1211+143 by
Pounds et al. (2003) that are distinct in the model spectra. The black
tick marks indicate lines identified in the observations that do
not have significant strength in any of the model spectra. 
Fe~{\sc xxiv} (blue tick mark) is responsible for the feature are 11~\AA~ in the $\mu =
0.475$ spectrum which was not discussed by Pounds et al. (2003).
}
\label{fig:pg1211_wide}
\end{figure}

The spectral model used here is too simplistic to address the strong soft
excess observed in PG1211+143.
However, we note that the wind model does clearly lead to an excess of
emission below $\sim 1$~keV suggesting that reprocessing by the fast
flow may account for a modest fraction of the excess soft-energy emission 
(c.f. \citealt{king03}). Much of this contribution is due to heavily
blended forests of emission lines (including those of L-shell Fe) but
there are also moderately strong discreet features, such as 
the broad O~{\sc viii} Ly$\alpha$ emission
line mentioned above.

We defer more detailed studies including both a wider
exploration of the model parameter space and the interplay of
wind-formed spectral features with other components of the system to
later work but the simple comparisons presented here already support
the notion that the Fe~K absorption {\it and} emission features could be
predominantly formed in a fast outflow as in the picture presented by
\cite{pounds09}. Scaled to the black hole mass of PG1211+143 ($4
\times 10^7$~M$_{\odot}$, \citealt{kaspi00}; see
Section~\ref{sect_rescale}), the model with which we have
compared has a mass-loss
rate of $\dot{M} = 2.1$~M$_{\odot}$~yr$^{-1}$ and a total covering
fraction\footnote{Although our wind has a total covering
  fraction of $b = 0.7$, this value cannot be directly compared
  to that obtained from the much simpler model of \cite{pounds09} 
  -- there is significant diversity in the line-of-sight properties
  of our model and some of the lines-of-sight through the wind have
  only very low optical depth.} of $b = 0.7$. These are comparable to the properties inferred
by \cite{pounds09} and support their conclusion that the flow is very
likely to be energetically important.

\section{Conclusions and future work}
\label{sect_conc}

Outflows from the accretion disks around supermassive black holes are
a promising explanation for the blueshifted absorption line features
that have been identified in the X-ray spectra of AGN. 
They are also theoretically expected in high Eddington ratio sources for
which line-driven winds 
have been modelled by e.g. \citep{proga04}.
AGN outflows may be massive enough to have significant implications for our
understanding of accretion by supermassive black holes 
and likely have implications for the interpretation of a wide range of both absorption and emission features
in the high energy spectra of AGN \citep{turner09}.
However, to
properly interpret the observed spectroscopic features,
modelling based on realistic theoretical spectra for outflow models is
required in order to quantify the flow properties.
Here, we have made a significant step towards this goal by 
extended the code described in Paper~I to incorporate the
physics of L- and M-shell ions and to
compute both the ionization and thermal state of the outflow in
detail. This allows us to explore a wider range of
plausible outflow conditions including, in particular, less
highly-ionized and more optically thick flows.

An example calculation for one of our simply-parameterized wind
models illustrates that, for mass-loss rates comparable to the
Eddington accretion rate, a wide range of physical conditions are
likely to be present in a fast outflow such that a very diverse range
of spectral signatures are possible. As in Paper~I, we find that considerable
complexity can be introduced to the Fe~K$\alpha$ region where both
narrow absorption and broad, red-skewed emission lines are predicted to
form. However, such an outflow can also significantly affect the
spectrum at softer energies -- narrow, highly blueshifted 
absorption lines of lighter
elements and lower ionization states of Fe are to be expected for many
line-of-sight through an outflow. 
Emission features associated with these lines also form 
which can be 
broadened and skewed in a manner similar to the Fe~K$\alpha$ line,
potentially allowing the wind model to account for broad Fe~L features
as have been reported in the spectrum of 1H0707-495
(see e.g. \citealt{fabian09,zoghbi09}).
These features in the soft band are
generally weaker that those in the Fe~K$\alpha$ region and their
interpretation is more likely to be complicated owing to the wide range
of physical components that may affect the softer bands (e.g. the
soft excess and/or warm absorbers).

As proof-of-concept, we have presented a simple comparison of
our outflow model spectra with the well-known quasar PG1211+143. We
found that the most important features that have been identified in
its hard X-ray spectra, namely a strong absorption line at
$\sim 7$~keV and a broad emission feature peaking around $6$~keV, can be
simultaneously well-matched by our theoretical spectra. Although we
have not fully explored the possible parameter-space of outflow models
nor the interplay with other physically motivated
phenomena (such as disk reflection), our results support the
interpretation of \cite{pounds09} that the absorption and emission
components are likely physically related and form in a fast outflow
that has a substantial covering fraction and a
mass-loss rate comparable to the Eddington accretion rate.

Our next step will be to attempt detailed model fits to one or more
AGN using our radiative transfer code. This will allow us to more
fully test the outflow paradigm and quantify the range of
outflow properties which are suggested by observations, a critical step in
establishing the role of highly-ionized outflows in the AGN phenomenon.

\section*{Acknowledgments}

SAS thanks Caroline D'Angelo for many useful discussions and
helpful suggestions.
TJT acknowledges NASA Grant NNX09AO92G.
We thank the anonymous referee for several constructive comments.

\bibliographystyle{mn2e}
\bibliography{snoc}

\label{lastpage}

\end{document}